%% file: sample.tex
\documentclass[sigconf]{acmart}

\input{preamble}

\usepackage{graphicx}
\usepackage{multirow}
\usepackage{booktabs}
\usepackage{array}
\usepackage{cellspace}  
\usepackage{rotating}
\usepackage{tabularx}
\usepackage{longtable}
\usepackage{hyperref}
\usepackage{amsmath} 
\usepackage{booktabs} 
\usepackage[dvipsnames]{xcolor}

\renewcommand{\arraystretch}{1.5}  

\begin{document}


\title[\revisesecond{Vision-Based Multimodal Interfaces: A Survey and Taxonomy for Enhanced Context-Aware System Design}]{\revisesecond{Vision-Based Multimodal Interfaces: A Survey and Taxonomy for Enhanced Context-Aware System Design}}

\author{Yongquan `Owen' Hu}
\orcid{0000-0003-1315-8969}
\affiliation{%
 \institution{School of Computer Science and Engineering,}
 \institution{University of New South Wales}
\city{Sydney}
\country{Australia}
}
\email{yongquan.hu@unsw.edu.au}

\author{Jingyu Tang} 
\orcid{0009-0008-8259-8531} 
\affiliation{%
\institution{School of Computer Science and Technology,} 
\institution{Huazhong University of Science and Technology} 
\city{Wuhan}
\country{China}
}
\email{u202215423@hust.edu.cn}
\authornote{Jingyu Tang and Xinya Gong contributed equally to this work.}

\author{Xinya Gong} 
\orcid{0009-0005-6414-9351}
\affiliation{%
 \institution{Department of Computer Science and Engineering}
 \institution{South University of Science and Technology}
 \city{Shenzhen}
 \country{China}
}
\email{gongxinya123@gmail.com}
\authornotemark[1] 

\author{Zhongyi Zhou} 
\orcid{0000-0002-1363-7313}  
\affiliation{%
\institution{The University of Tokyo}
\city{Tokyo}
\country{Japan}
}
\email{zhongyi.zhou.work@gmail.com}

\author{Shuning Zhang}
\orcid{0000-0002-4145-117X}
\affiliation{%
\institution{Tsinghua University}
\city{Beijing}
\country{China}
}
\email{zsn23@mails.tsinghua.edu.cn}

\author{Don Samitha Elvitigala}
\orcid{0000-0002-8013-5989}
\affiliation{%
\institution{Exertion Games Lab, Department of Human-Centred Computing,}
\institution{Monash University}
\city{Melbourne}
\country{Australia}
}
\email{don.elvitigala@monash.edu}

\author{Florian `Floyd' Mueller} 
\orcid{0000-0001-6472-3476}
\affiliation{%
\institution{Exertion Games Lab, Department of Human-Centred Computing,}
\institution{Monash University}
\city{Melbourne}
\country{Australia} 
}
\email{floyd@exertiongameslab.org}

\author{Wen Hu}
\orcid{0000-0002-4076-1811}
\affiliation{%
\institution{School of Computer Science and Engineering,}
\institution{University of New South Wales}
\city{Sydney}
\country{Australia}
}
\email{wen.hu@unsw.edu.au}

\author{Aaron J. Quigley}
\orcid{0000-0002-5274-6889}
\affiliation{%
\institution{CSIRO's Data61 \& }
\institution{University of New South Wales}
\city{Sydney}
\country{Australia}
}
\email{aquigley@acm.org}

\renewcommand{\shortauthors}{Hu et al.}

\begin{abstract}
The recent surge in artificial intelligence, particularly in multimodal processing technology, has advanced human-computer interaction, by altering how intelligent systems perceive, understand, and respond to contextual information (i.e., context awareness). Despite such advancements, there is a significant gap in comprehensive reviews examining these advances, especially from a multimodal data perspective, which is crucial for refining system design. This paper addresses a key aspect of this gap by conducting a systematic survey of data modality-driven Vision-based Multimodal Interfaces (VMIs). VMIs are essential for integrating multimodal data, enabling more precise interpretation of user intentions and complex interactions across physical and digital environments. Unlike previous task- or scenario-driven surveys, this study highlights the critical role of the visual modality in processing contextual information and facilitating multimodal interaction. 
\revise{
Adopting a design framework moving from the whole to the details and back, it classifies VMIs across dimensions, providing insights for developing effective, context-aware systems.}
\end{abstract}




\begin{teaserfigure}
  \centering
  \includegraphics[width=0.88\textwidth]{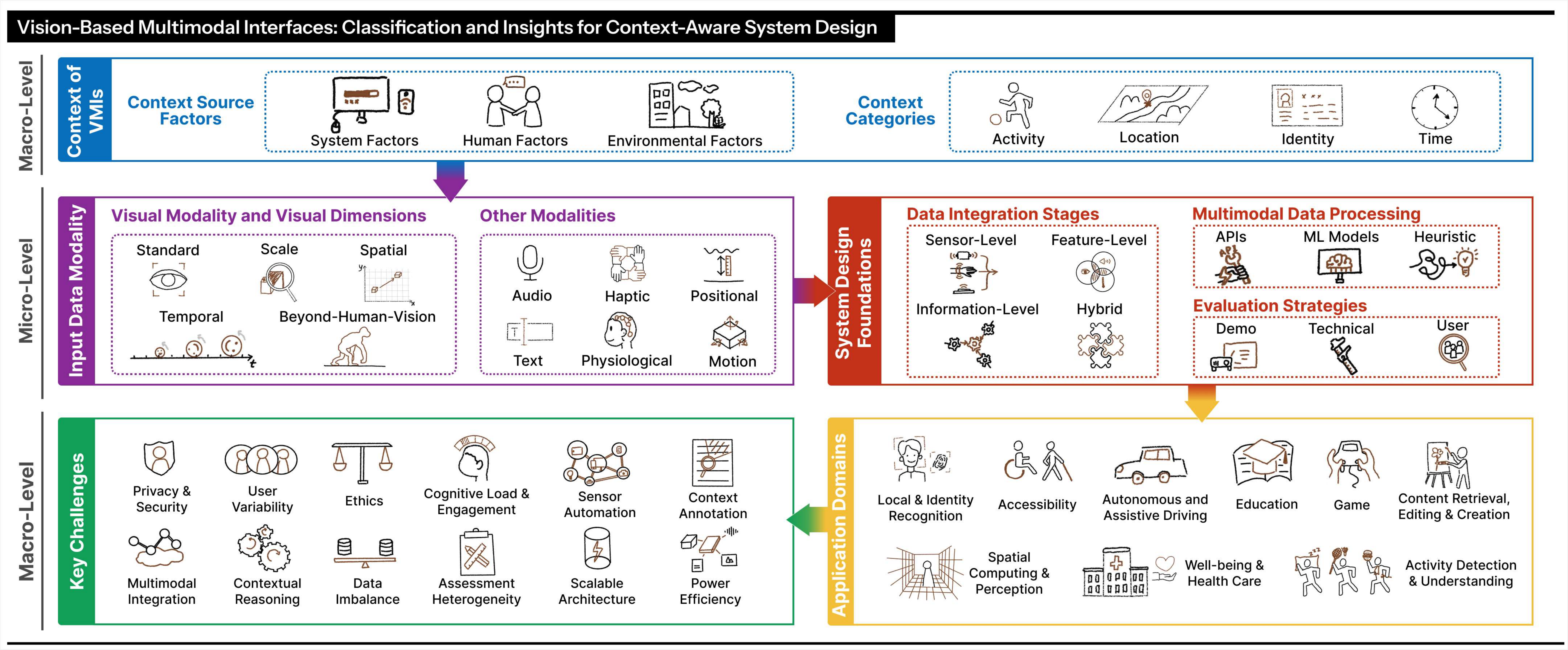}
  \caption{\revisesecond{We review and categorize VMIs aimed at enhancing context awareness. Our key contribution is a \textbf{\textit{Macro-Micro-Macro level}} (whole-detail-whole) system design framework, providing actionable references from a \textbf{\textit{Data Modality-Driven}} perspective:
(1) Macro-level contextual factors: considerations for context understanding (Section~\ref{taxonomy_context});
(2) Micro-level system foundations: input data modality (visual + other modalities), data integration stages, multimodal data processing and evaluation strategies (Sections~\ref{sec: data modality},~\ref{sec: system_design_foundations});
(3) Macro-level design synthesis: application domains, design considerations and key challenges (Sections~\ref{sec: application_domains},~\ref{sec: design_challenges}).}}
  \label{fig:teaser}
\end{teaserfigure}

\maketitle

\section{Introduction}

Context awareness is essential in Human-Computer Interaction (HCI), enabling systems to detect, interpret, and respond to contextual information~\cite{schilit1994context, Abowd1999TowardsAB, harter1999anatomy}, thereby facilitating adaptive and seamless interactions~\cite{weiser1999computer}. \revise{Vision-based interfaces (VIs), such as camera-based gesture recognition~\cite{starner1995visual, hasan2012human}, excel in interpreting complex visual data for tasks like fine-grained gesture recognition or spatial context analysis~\cite{kolsch2004vision, hu2024towards}. VIs also enable unobtrusive interactions, supporting calm computing by reducing cognitive load~\cite{schmidt2000implicit, dey2005designing, weiser1996designing}. Applications include smart homes and immersive environments such as Virtual Reality (VR) and Augmented Reality (AR)~\cite{makkonen2009context,haque2020illuminating,ismail2015vision}. Vision-based Multimodal Interfaces (VMIs) enhance context awareness by integrating visual inputs with non-visual modalities, creating a unified understanding of the environment~\cite{oviatt2007multimodal,shin2007vision}. For instance, VEmotion combines visual, GPS, and auditory data to improve driver emotion recognition accuracy by 28.5\% compared to visual-only approaches~\cite{Bethge2021VEmotionUD}, demonstrating how VMIs address unimodal limitations to deliver accurate, contextually relevant responses~\cite{sharma1998toward,ismail2015vision}.}

\begin{figure*}
    \centering
    \includegraphics[width=0.8\linewidth]{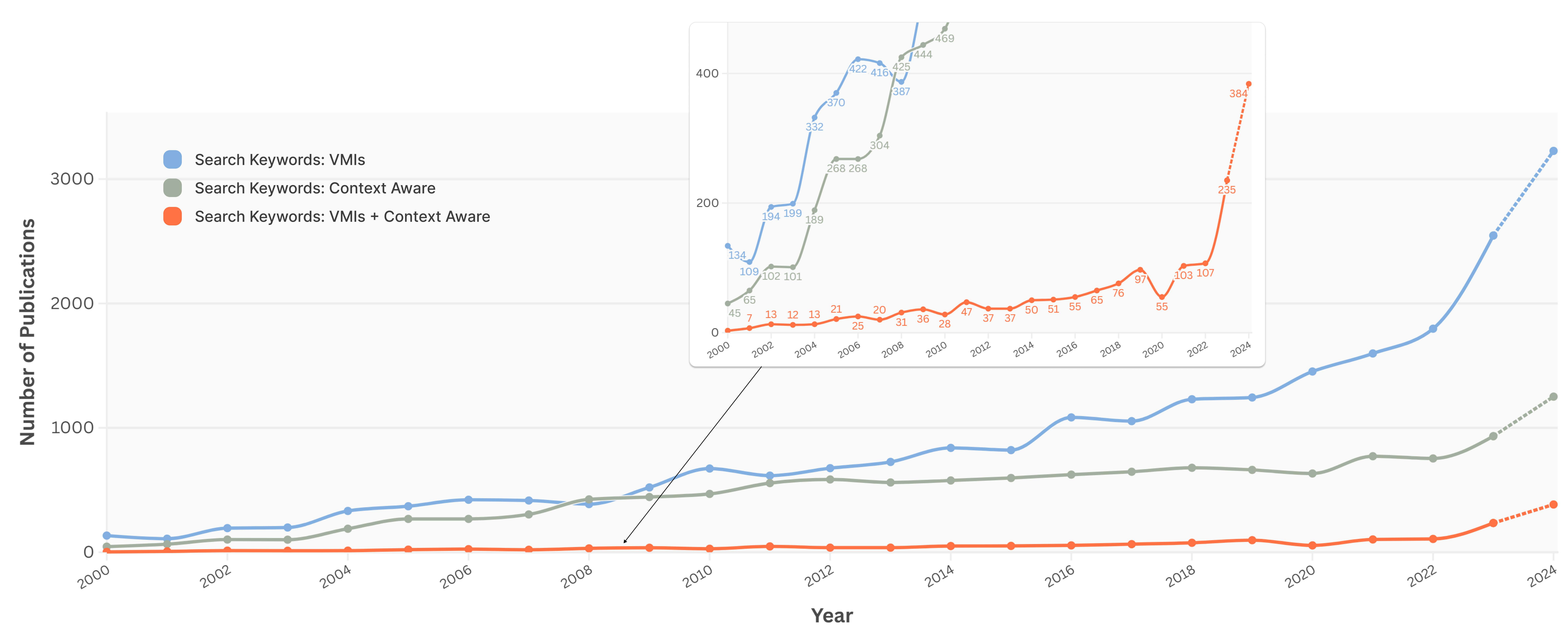}
    \caption{\revise{The publication growth trend for vision-based multimodal interface and context awareness in the ACM Digital Library (details of the search keywords are provided in Appendix~\ref{select_literature}).}}
    \label{fig:trend}
\end{figure*}

Despite ongoing advancements in visual information processing, the use of VMIs as interactive tools remains nascent. The rapid development of technology, especially in multimodal Artificial Intelligence (AI), has outpaced the design principles and paradigms of interactive systems, creating a gap in intuitive systems that can be easily utilized by non-experts~\cite{yildirim2024multimodal,schuller2021towards}. As a result, VMIs have gained increasing attention in the HCI community for their ability to address the challenges of integrating multimodal data and enabling effective interactions within this evolving paradigm. As shown in Figure~\ref{fig:trend}, there is a growing recognition of the value of research in this area, as evidenced by the increasing volume of related work (details in Section~\ref{literature_selection_methodology}). Significantly, the expanding intersection of VMIs with context awareness reflects an emerging trend toward integrating these interfaces with a deeper understanding of the contextual factors influencing user interactions. 
\revise{
Our study builds upon previous research, aligning with the growing interest in integrating multimodal data for context-aware systems, and provides a thorough and up-to-date analysis with practical guidance and systematic insights for advancing VMI design.
}

There exist related VMI-related surveys, particularly focusing on their applications in fields where emerging technologies enhance system capabilities, such as augmented reality enabling more immersive experiences~\cite{suzuki2022augmented}, Generative AI (GenAI) improving content personalization, or mobile computing facilitating real-time interactions in dynamic environments~\cite{hu2023exploring}. 
\revise{In practice, building VMIs requires systematic efforts and there is a lack of actionable frameworks with practical insights in the community to guide the HCI practitioners to prototype interactive systems. For example, while Dumas et al.~\cite{dumas2009multimodal} provide theoretical foundations for multimodal interfaces, they offer limited practical guidance for complex, context-aware scenarios. This type of practical guidance, which is currently lacking in the field, is essential for bridging the gap between theory and practice and offering clear pathways for designing effective context-aware systems.}
\revise{Additionally, many studies emphasize specific perspectives, such as task-driven (e.g., gesture recognition~\cite{hasan2012human}), scenario-driven (e.g., mobile and AR~\cite{kolsch2004vision,suzuki2022augmented,Grubert2017TowardsPA}), or technology-driven (e.g., GenAI~\cite{shi2023hci}) approaches, which could inadvertently overlook the potential of a broader data perspective. These perspectives provide unified frameworks for diverse modalities, addressing dynamic real-world environments, characterized by rapidly changing contexts, user behaviors, and system requirements~\cite{sezer2017context}.}
For instance, Bolchini et al.~\cite{bolchini2007data} demonstrated how data tailoring enhances adaptability, such as adjusting museum guides for low-vision users or personalizing content based on user interests. \revise{It underscores the value of a data-oriented approach in enabling systems to flexibly combine and adapt information for more effective and personalized interactions.}
Building on this foundation, our study adopts a \textbf{\textit{data modality-driven}} lens to evaluate strategies for integrating visual modalities with other data streams, explore innovative methods, and assess their roles in enhancing interactive experiences within context-aware systems.

\revise{
Our study offers a systematic review, organizing VMIs into a taxonomy using a \textbf{\textit{Macro-Micro-Macro (3M) level}} system design framework, which seeks to address the aforementioned limitations.
This framework aligns with the top-down and bottom-up design philosophy~\cite{borenstein2004combining,dix2004human} of moving from the holistic perspective (Section~\ref{taxonomy_context}) to finer details  (Section~\ref{sec: data modality},~\ref{sec: system_design_foundations}) and then synthesizing them back into an integrated whole (Section~\ref{sec: application_domains},~\ref{sec: design_challenges}), as shown in Figure~\ref{fig:teaser}.
}
\revise{This structure is intended not only to bridge existing gaps in usability but also to serve as an iterative \textit{\textbf{step-by-step manual organized by sections}} for practitioners. For instance, we visualizes information flow across dimensions with the help of \textit{\textbf{a Sankey diagram}} and \textit{\textbf{an interactive website}}, while Appendix~\ref{literature_detail} provides \textit{\textbf{detailed literature statistics} (categories, counts, and citations) to enhance practical usability.}}
\revise{Moreover, by adopting a data modality-driven perspective, our study highlights its adaptability and flexibility in integrating diverse data modalities. Unlike task-driven or scenario-driven perspectives, this approach provides a systematic framework for addressing challenges such as cross-modal semantic alignment and data integration. For instance, Cai et al.~\cite{cai2019survey} demonstrate how integrating multimodal data streams in smart healthcare systems enhances diagnostic accuracy and supports robust decision-making processes. These findings suggest that a data modality-driven perspective informs VMI design by structuring visual and non-visual modality integration, especially in complex scenarios. This perspective bridges theoretical insights with practical applications, enabling more adaptive and scalable interaction designs.}

\section{Scope and Methodology}\label{methodology}

\subsection{Scope and Definitions}
In this section, we aim to establish a clear scope and definitions of the terminology used throughout this paper.

\subsubsection{Context Awareness and Visual Data}
\revise{Context awareness is a foundational concept in HCI, enabling systems to perceive and respond to environmental changes dynamically~\cite{makkonen2009context}. Initially focused on static factors like user location, it has evolved to encompass dynamic properties shaped by interactions and activities~\cite{schilit1994context,dey2001conceptual,dourish2004we}, supporting adaptive interfaces, personalized data, and smart environments~\cite{bolchini2007data}. Visual data has been pivotal in advancing these capabilities, as demonstrated by Schilit et al.'s navigation systems using visual feedback~\cite{schilit1994context} and Dey and Abowd’s applications for real-time tracking and location-based guidance~\cite{dey2001conceptual,dey2005designing}. Over time, the role of visual data expanded through integration with other modalities in VMIs, enabling real-time spatial and gesture recognition in XR systems~\cite{benko2008sphere,surale2019tabletinvr}. These multimodal approaches allow systems to interpret complex contexts, such as subtle gestures or dynamic environments, and tailor interactions to diverse users, enhancing accessibility~\cite{peissner2012myui,he2023multi}. To support these advancements, established frameworks like Dey and Abowd’s~\cite{Abowd1999TowardsAB} categorize context into location, identity, activity, and time, while Grubert et al. emphasize high-level factors like human and environmental elements~\cite{Grubert2017TowardsPA}. Rather than creating a new framework, our work adapts these taxonomies to VMIs with application-specific customizations, preserving their strengths while tailoring them to the unique requirements of multimodal systems. This approach bridges theoretical foundations with practical applications, making visual data a cornerstone of adaptable and effective context-aware technologies.}

\begin{figure*}
    \centering
    \includegraphics[width=\linewidth]{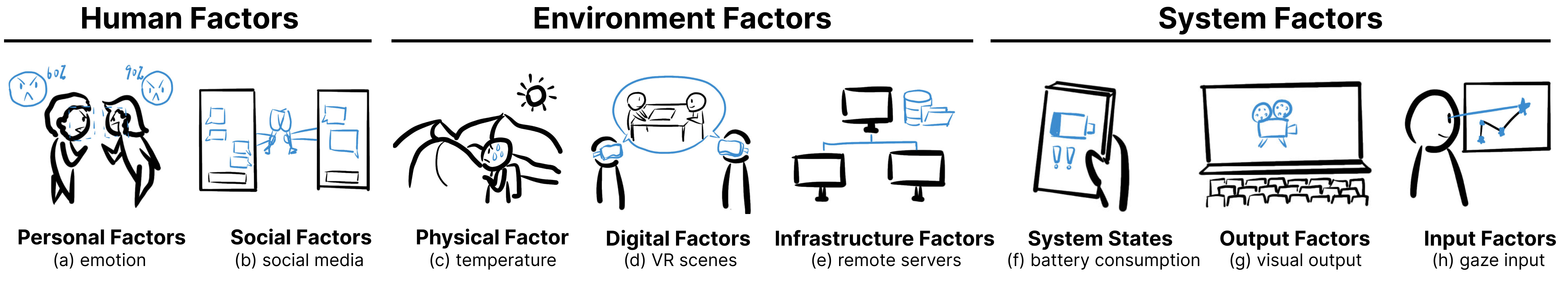}
    \caption{\revise{Examples of context source factors in VMIs with descriptions and citations (illustrative references (a):~\cite{Bethge2021VEmotionUD}, (b):~\cite{Rawat2017ClickSmartAC}, (c):~\cite{Saad2023HotFootFU}, (d):~\cite{Chen2023PaperToPlaceTI}, (e):~\cite{Wang2024GVOILAGI}, (f):~\cite{Dogan2021SensiCutML}, (g):~\cite{Chen2023PaperToPlaceTI}, (h):~\cite{Lee2024GazePointARAC}).}}
    \label{fig:context_source_factor}
\end{figure*}

\subsubsection{Vision-Based Interfaces}

\revise{
We define a VI by synthesizing insights from various sources: Kuno et al.~\cite{kuno1994vision} highlight the intuitive potential of visual input for user-centered interactions without rigid calibration; Sá et al.~\cite{sa2001vision} demonstrate its integration with other modalities to enhance contextual awareness; Zabulis et al.~\cite{zabulis2009vision} and Gopalan et al.~\cite{gopalan2009toward} focus on processing visual input for capturing user behaviors and environmental factors through gesture recognition and posture analysis; and Kolsch et al.~\cite{kolsch2004vision} emphasize system adaptability in dynamic environments. Building on these contributions, we define a VI as \textit{a system that utilizes \textbf{visual input from one or more sensors (e.g., cameras)}, capturing user-relevant factors (e.g., human behavior, environment) to facilitate human-machine interaction.} While the literature on VIs shares commonalities, it varies in descriptions that we exclude from our definition, considering them as features rather than essential criteria. For instance, real-time processing, though important for enhancing interaction~\cite{kolsch2004vision,hua2022vision,zabulis2009vision}, is not mandatory, as some studies prioritize algorithm accuracy over real-time capabilities~\cite{Suzuki2019AnOM}. Similarly, while most studies have pointed out cameras as primary sensors, some briefly mention~\cite{Suzuki2019AnOM,Zhou2022ContextAware3O,Fan2024ContextCamBC,Yang2023ContextAwareTV} or exclude them entirely~\cite{Lian2023MultitaskLF}.
}

\subsubsection{Vision-based Multimodal Interface and Enhanced Context-Aware System} \label{vmis_definition}
Our definition of the VMI builds on the foundational concept of a VI, incorporating insights from related literature~\cite{tong2024cambrian, shin2007vision, sharma1998toward}. Specifically, a VMI is \textit{a subset of VI, characterized by the inclusion of \textbf{at least one non-visual modality in addition to the visual modality or a combination of two or more distinct visual dimensions within its input data}} (detailed in Section~\ref{sec: data modality}). This information, related to the user, is used to enhance interaction. VMIs can also enhance context awareness by integrating multimodal data, enabling systems to capture context more accurately~\cite{kernchen2005multimodal}. While Salber noted the differences between multimodal systems, which rely on explicit input, and context-aware systems, which use implicit input~\cite{salber2000context}, research has expanded multimodal systems to include both input types~\cite{ruiz2009multimodal}. 
\revise{
In this paper, we focus on the intersection of these definitions, with VMIs incorporating both explicit and implicit inputs. Notably, visual input in VMIs often functions as an implicit source of context-awareness, highlighting its critical role in understanding and adapting to dynamic environments.
}

\subsection{Contributions}\label{contributions}
\revise{
In this paper, we make three key contributions. First, we \textit{\textbf{systematically review the literature}} across HCI venues, identifying trends and underexplored areas through a \textit{\textbf{data modality-driven perspective}}. We place a particular focus on the visual modality, while integrating data from other modalities to address the dynamic requirements of context awareness. Second, we propose a taxonomy structured within the \textit{\textbf{3M framework}} for system design, providing an \textit{\textbf{actionable reference}} that includes an iterative process and practical resources, such as \textit{\textbf{interactive website}}, to guide the development of context-aware systems. Third, we identify \textit{\textbf{key design considerations}} and \textit{\textbf{open research challenges}}, offering future directions for advancing VMIs and multimodal interaction paradigms.
}

\subsection{Literature Selection Methodology}\label{literature_selection_methodology}

\subsubsection{Literature Search and Selection}
\revise{
We conducted a systematic literature search in digital libraries including ACM and IEEE, following the PRISMA framework~\cite{PRISMAStatement}. Using the query \texttt{("vision-based") AND ("multimodal") AND ("context aware")} and related synonyms, we targeted English-language publications since 2018.} This search was informed by factors such as research trends, advancements in the field, and paper volume. After removing duplicates, 929 papers remained, which were reviewed to exclude works outside the scope of our study, such as single-modality interfaces or non-HCI-related literature. This process resulted in 98 relevant papers. To complement the search, expert discussions added 11 significant works, yielding a curated collection of 109 papers.
\revise{Further details on the search process and selection criteria are provided in  Appendix~\ref{select_literature}.}

\subsubsection{Analysis and Synthesis}
The dataset was analyzed through a multi-step process. First, we conducted open coding on a small subset of our sample to identify an initial approximation of the dimensions and categories within the design space. Next, we reviewed the initial classification to assess the consistency and comprehensiveness of the categorization methods, during which categories were merged, expanded, or removed. Following this, we systematically coded the entire dataset, applying individual tags for precise categorization. Finally, we reviewed the individual tags to resolve any discrepancies and arrive at the final coding results.
\revise{To minimize bias, ensure comprehensive assessment, and enhance reliability and transparency, the data was independently coded and analyzed by four co-authors, with the results subsequently consolidated~\cite{PRISMAStatement}.}

\section{Context of VMIs} \label{taxonomy_context}
\revise{
In this section, we build on previous research to refine context classification, highlighting the ultimate goal of system design as its whole guiding factor.
}

\begin{figure*}
    \centering
    \includegraphics[width=0.7\linewidth]{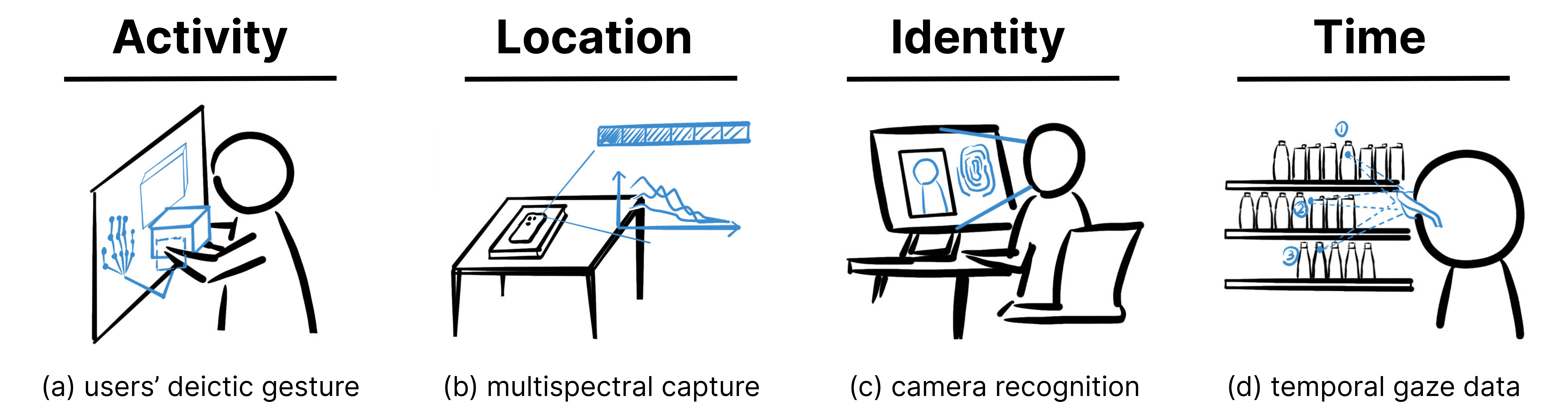}
    \caption{\revise{Examples of context categories in VMIs with descriptions and citations (illustrative references (a):~\cite{zhou2022gesture}, (b):~\cite{Yeo2017SpeCamSS}, (c):~\cite{Liang2021AuthTrackEA}, (d):~\cite{Lee2024GazePointARAC}).}}
    \label{fig:context_categories}
\end{figure*}

\subsection{Context Source Factors} \label{context_source_factors}
To gain a deeper understanding of the factors influencing context, we build on the classification of situations proposed by Grubert et al.~\cite{Grubert2017TowardsPA}. However, we extend its application beyond AR scenarios to encompass a broader range of fields. Additionally, we refine their three major categories—human factors, environmental factors, and system factors—by introducing a more fine-grained classification, accompanied by examples for clarification and illustration, as shown in Figure~\ref{fig:context_source_factor}.

\textcolor{NavyBlue}{\textbf{\textit{Factor-1. Human Factors:}}}
Classically, \emph{Human Factors} refer to the study of how humans interact with elements of a system or environment. This field focuses on understanding and improving how well people interact with the systems they use. Human factors can be categorized into personal and social factors.

\textcolor{NavyBlue}{\textit{-Personal Factors:}} Focusing on an individual user, personal factors such as cognitive load~\cite{Wen2024AdaptiveVoiceCA, Cai2024PANDALensTA, Zhu2023IntegratingGA}, emotion~\cite{Fan2024ContextCamBC, Cai2024PANDALensTA, Bethge2021VEmotionUD, Meng2022ValenceAA} and user preference~\cite{Cai2024PANDALensTA, Wen2024AdaptiveVoiceCA, Chen2023PaperToPlaceTI, xu2024can} are frequently considered in interface design to promote a tailored user experience. Furthermore, user posture and physical movement~\cite{Wang2024PepperPoseFP, Wang2023UbiPhysio, Liang2021AuthTrackEA, Ahuja2021PoseontheGoAU} constitute significant personal factors, underscoring the need for adaptive design strategies.

\textcolor{NavyBlue}{\textit{-Social Factors:}} In contrast to personal factors, social factors encapsulate the interactions and relational networks between multiple individuals. These elements include, but are not limited to, social media information~\cite{Rawat2017ClickSmartAC}, as well as social norms and social zones~\cite{Schwrer2023Nav2CANAC}. Together, these factors provide a richer understanding of social interplay, thereby improving the design and effectiveness of systems.

\textcolor{NavyBlue}{\textbf{\textit{Factor-2. Environment Factors:}}}
\emph{Environment Factors} describe the surrounding of the user and the interfaces in which interaction takes place. Within the domain of environmental factors, we distinguish between physical factors, digital factors and infrastructure factors. 

\textcolor{NavyBlue}{\textit{-Physical Factors:}} Physical factors encompass environmental elements of the physical world. Raw factors, such as temperature~\cite{Wang2020CAPturARAA, Saad2023HotFootFU, Zhao2022MediatedAT}, light levels~\cite{Zhao2022MediatedAT, Yeo2017SpeCamSS, Lim2024ExploringCM, schrapel2021spectrophone}, and noise levels~\cite{Lim2024ExploringCM}, can be directly perceived by human senses or measured via sensors.
 \revise{Derived factors, by contrast, are calculated by combining multiple raw factors or abstracting higher-level information from low-level data. For example, the spatial or geometric configuration of a scene can be inferred from multispectral or visual data~\cite{Fender2018VeltAF, Sun2023TemporallyCS, Yang2012MagicFA, Yeo2017SpeCamSS, schrapel2021spectrophone}. Similarly, the presence or absence of physical artifacts, such as objects or materials, can provide contextual insights~\cite{Deng2023ContextAwareFF, Lin2014ACA, Lian2023MultitaskLF}.}
 

\textcolor{NavyBlue}{\textit{-Digital Factors:}} In contrast to physical factors, the second category of environmental factors focuses on the digital environment. This category encompasses information stored or displayed in virtual environments such as conversation logs~\cite{xu2024can}, online conference data~\cite{Hautasaari2024EmoScribeCA} and VR/AR scenes~\cite{Wen2024AdaptiveVoiceCA, Chen2023PaperToPlaceTI, Gupta2023SensoryScapeCE, Suzuki2019AnOM}.

\textcolor{NavyBlue}{\textit{-Infrastructure Factors:}} Unlike human and physical factors, which are often subject to immediate perception and direct manipulation, infrastructure factors operate behind the scenes, subtly yet powerfully influencing the system's dynamics. This category encompasses elements including network connectivity~\cite{Yeo2017SpeCamSS, Cho2018DeepTI, Yang2012MagicFA, Wen2024AdaptiveVoiceCA, Wang2024GVOILAGI, xu2024can, Abbas2018AHA, Liang2021AuthTrackEA, Ahuja2021PoseontheGoAU,Chen2021NeckFace}, databases~\cite{Jamonnak2021GeoContextAS, Abbas2018AHA} and remote server access~\cite{Fan2024ContextCamBC, Wang2024GVOILAGI, McDuff2019AME}.

\textcolor{NavyBlue}{\textbf{\textit{Factor-3. System Factors:}}}
These include general system configuration, computational capabilities of the device, output and input devices, and modalities. System factors can be classified into system state, output factors, and input factors.

\textcolor{NavyBlue}{\textit{-System States:}} the system state refers to the current availability and performance of a system's computational resources, including factors such as memory usage~\cite{Malawade2022HydraFusionCS}, latency~\cite{Wang2023UbiPhysio, Malawade2022HydraFusionCS}, and battery consumption~\cite{Dogan2021SensiCutML, hu2023microcam, xu2024can, meyer2022u, Meyer2021ACH, Malawade2022HydraFusionCS, schrapel2021spectrophone}. These elements collectively determine the system's capacity to handle tasks and its operational efficiency at any given time.


\textcolor{NavyBlue}{\textit{-Output Factors:}} Output factors refer to the various ways information is presented to the user, including visual output~\cite{Fan2024ContextCamBC, Chen2023PaperToPlaceTI, Hwang2020MonoEyeMH, Gupta2023SensoryScapeCE, Fender2018VeltAF, Lian2023MultitaskLF, Suzuki2019AnOM}, as well as other modalities like audio~\cite{Wen2024AdaptiveVoiceCA, Lim2024ExploringCM, Su2024SonifyARCS, Khan2021PALWA, Lin2014ACA} or haptic feedback~\cite{Li2021TowardsCA,chen2024video2haptics}.

\textcolor{NavyBlue}{\textit{-Input Factors:}} Input factors describe the different methods available for users to interact with the system, such as gestures~\cite{Fleer2012MISOAC, Lee2024GazePointARAC}, haptic input~\cite{Athar2023VisTacTA, Ahuja2021PoseontheGoAU, wang2021elastic}, mouse input~\cite{Zhu2023IntegratingGA}, gaze~\cite{Lee2024GazePointARAC, Wen2024AdaptiveVoiceCA, Matsuda2018EmoTourME, Chen2023PaperToPlaceTI, Wang2024GVOILAGI, meyer2022u}, or speech~\cite{Wang2024WatchYM, Cai2024PANDALensTA, Bethge2021VEmotionUD, Kianpisheh2024exHAR, Wang2024GVOILAGI}. Depending on the input modalities available, the system can adapt its operation to best suit the user's input method.

\subsection{Context Categories} \label{context_categories}
Classifying context types is crucial for application designers to identify the most relevant aspects of context for their specific applications. Although technological advancements and changing application scenarios have led to the evolution of context classification, the general framework~\cite{Abowd1999TowardsAB} remains a foundational reference for many studies. In this section, we adhere to their classic four-classification method (i.e., \textit{activity}, \textit{location}, \textit{identity} and \textit{time}) and provide examples to illustrate the role of VMIs in each category in Figure~\ref{fig:context_categories}.

\textcolor{NavyBlue}{\textbf{\textit{Category-1. Activity:}}}
VMIs in activity contexts leverage wearable technologies to recognize user actions like gestures and postures, enabling intuitive interactions. Key applications include vision-based gesture recognition in mixed reality and object annotation~\cite{Lee2024GazePointARAC, zhou2022gesture}, as well as gesture-based commands for robots, enhancing intent interpretation~\cite{Fleer2012MISOAC, Uimonen2023AGM, kong2021eyemu}. Gaze detection, another major use, tracks attention and assesses cognitive load using eye-tracking and neural networks~\cite{Lee2024GazePointARAC, Wen2024AdaptiveVoiceCA, Wu2022ContextawareRD}. VMIs also analyze head movements, speech, and driving behaviors across diverse contexts like video conferencing and healthcare~\cite{Mittal2021MultimodalAC, Siddiqi2021AUA}. Ultimately, they enhance activity detection accuracy and expand interaction possibilities with multimodal data integration.

\textcolor{NavyBlue}{\textbf{\textit{Category-2. Location:}}}
VMIs for location sensing use visual data directly or combine it with other modalities. Multispectral imaging detects material placement, as shown in SpeCam and SpectroPhone~\cite{Yeo2017SpeCamSS, schrapel2021spectrophone}, while systems like MicroCam combine RGB and IMU data for enhanced accuracy~\cite{Yang2012MagicFA, hu2023microcam}. Non-visual methods like GPS, IMU, and radar enable faster detection; for instance, ContextCam integrates GPS and Wi-Fi~\cite{Bethge2021VEmotionUD}, while radar and Go-Pro data estimate user positions~\cite{elvitigala2023radarfoot}. Although visual approaches provide rich contextual information, non-visual methods offer lighter, faster solutions, often complemented by visual data for broader applications like facial recognition.

\textcolor{NavyBlue}{\textbf{\textit{Category-3. Identity:}}}
Identity recognition in VMIs answers ``Who is involved?'' and enhances contextual understanding. Auth+Track combines one-time authentication (e.g., iris, fingerprint) with continuous camera-based tracking to secure mobile use~\cite{Liang2021AuthTrackEA}. Khan et al.'s PAL integrates cameras and sensors for user authentication, supporting timely habit interventions~\cite{Khan2021PALWA}. For devices, AirConstellations and Kratos+ facilitate cross-device interactions and access control in multi-user environments~\cite{marquardt2021airconstellations, sikder2022s}. VMIs for user identity raise privacy concerns, while device identity is central to multi-device interactions.

\textcolor{NavyBlue}{\textbf{\textit{Category-4. Time:}}}
Time in VMIs determines when systems act and tag context for later retrieval~\cite{Abowd1999TowardsAB}. Real-time systems like GazePointAR and G-VOILA use gaze and gestures to ensure accurate, timely responses~\cite{Lee2024GazePointARAC, Wang2024GVOILAGI}. Temporal data also triggers actions, such as MicroCam capturing images based on phone placement~\cite{hu2023microcam}, or dynamically adjusting AR information based on user preferences~\cite{guarese2020augmented}. Whether in real-time or staged systems, temporal considerations are vital for effective context-aware interactions.

\section{Input Data Modality}\label{sec: data modality}

As outlined in Section~\ref{methodology}, VMIs are characterized by the modalities of their input data, which are typically acquired through sensing methods~\cite{dumas2009multimodal,gopalan2009toward}. To explore the multimodality of VMIs systematically, we begin with a detailed analysis of the visual modality, as it forms the core of our research focus. This analysis is further structured into several common dimensions, providing a foundation for a deeper understanding of VMIs from a multimodal data perspective. In addition, we briefly review other modalities, focusing on their data sources and functionalities, to offer a comprehensive view of how they complement the visual modality within multimodal systems.

\begin{table*}[h!]
\centering
\caption{\revise{Visual modality dimensions and corresponding examples of types and uses.}}
\label{tab:visual_example}
\resizebox{\textwidth}{!}{%
\begin{tabular}{m{4cm} m{5.5cm} m{5.5cm}} 
\Hline 
\textbf{Visual Modality Dimensions} & \multicolumn{2}{c}{\textbf{Examples of Types and Uses}} \\ 
\hline
\textbf{Standard-Vision} & 
\parbox[c][3.5cm][c]{5.5cm}{\raggedright RGB image for gaze capture~\cite{Ahuja2021PoseontheGoAU}\\ \includegraphics[width=2.5cm,height=2.5cm]{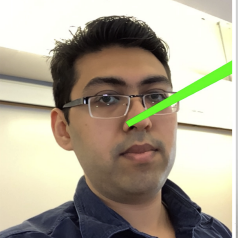}} & 
\parbox[c][3.5cm][c]{5.5cm}{\raggedright grayscale image of jacket texture~\cite{Yang2012MagicFA}\\ \includegraphics[width=2.5cm,height=2.5cm]{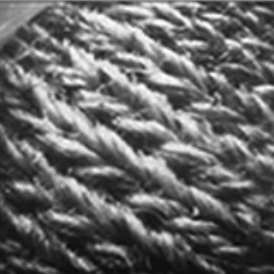}} \\ \hline
\textbf{Scale} & 
\parbox[c][3.5cm][c]{5.5cm}{\raggedright microscopic image of plush~\cite{hu2023microcam}\\ \includegraphics[width=2.5cm,height=2.5cm]{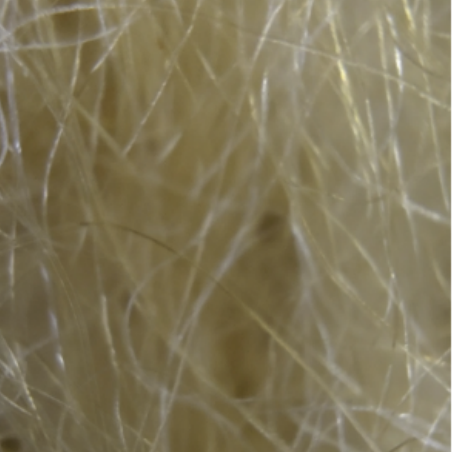}} & 
\parbox[c][3.5cm][c]{5.5cm}{\raggedright fisheye image for posture recognition~\cite{Liang2021AuthTrackEA}\\ \includegraphics[width=2.5cm,height=2.5cm]{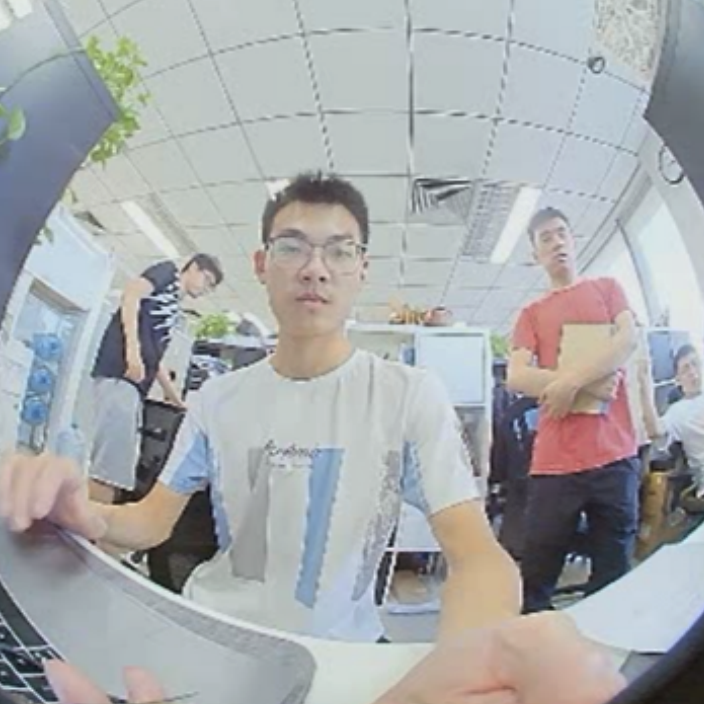}} \\ \hline
\textbf{Spatial} & 
\parbox[c][3.5cm][c]{5.5cm}{\raggedright depth image for face detection~\cite{Wang2024WatchYM}\\ \includegraphics[width=2.5cm,height=2.5cm]{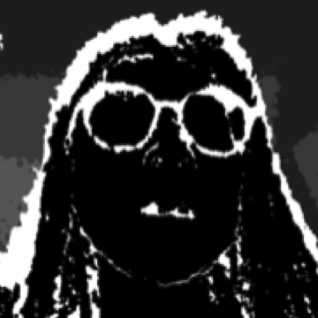}} & 
\parbox[c][3.5cm][c]{5.5cm}{\raggedright LiDAR image of room scaning~\cite{Deng2023ContextAwareFF}\\ \includegraphics[width=2.5cm,height=2.5cm]{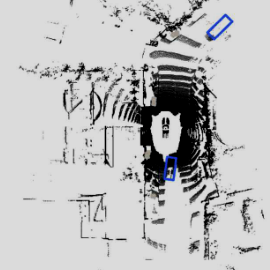}} \\ \hline
\textbf{Temporal} & 
\parbox[c][3.5cm][c]{5.5cm}{\raggedright event image for motion tracking~\cite{chen2024video2haptics}\\ \includegraphics[width=2.5cm,height=2.5cm]{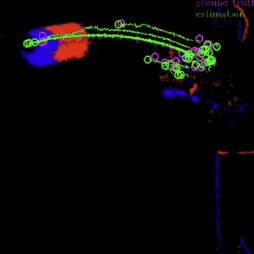}} & 
\parbox[c][3.5cm][c]{5.5cm}{\raggedright video for dynamic operations~\cite{liao2024realityeffects}\\ \includegraphics[width=2.5cm,height=2.5cm]{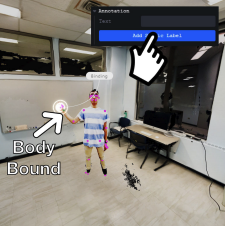}} \\ \hline
\textbf{Beyond-Human-Vision} & 
\parbox[c][3.5cm][c]{5.5cm}{\raggedright infrared image of human neck~\cite{Chen2021NeckFace}\\ \includegraphics[width=2.5cm,height=2.5cm]{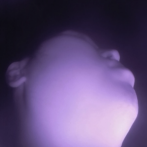}} & 
\parbox[c][3.5cm][c]{5.5cm}{\raggedright multispectral image of breadboard~\cite{Yeo2017SpeCamSS}\\ \includegraphics[width=2.5cm,height=2.5cm]{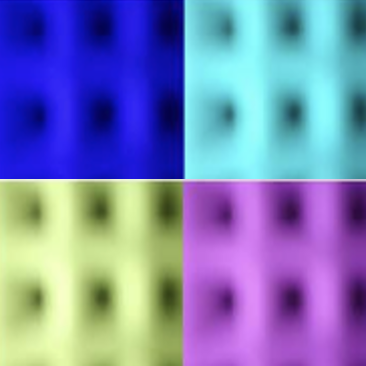}} \\ \Hline 
\end{tabular}
}
\end{table*}

\subsection{Visual Modality and Visual Dimensions}\label{vision_modality_dimension}
Images, typically captured by cameras, are the most common medium in the visual modality. Accordingly, we categorize the visual dimension based on image concepts.

\textcolor{Plum}{\textbf{\textit{Dimension-1. Standard-Vision:}}}
The Standard Vision dimension refers to standard visible images captured by cameras, primarily in the form of RGB or grayscale images. For example, Su et al. utilized RGB images from the rear camera of a mobile phone in RASSAR to recognize and reconstruct objects, facilitating seamless navigation indoors and outdoors~\cite{Su2024RASSARRA}. Similarly, Fan et al. employed the front camera of a mobile phone to detect user expressions using visible cues, enhancing the GenAI-based image creation process~\cite{Fan2024ContextCamBC}.
In VMIs, the Standard Vision dimension serves as the primary source of visual data, enabling systems to recognize objects, track movements, and detect interactions in real-time. Its ability to capture detailed and color-accurate visual information makes it indispensable for interpreting environmental cues.

\textcolor{Plum}{\textbf{\textit{Dimension-2. Scale:}}}
The scale dimension pertains to the scope or extent of the scene captured by the image, ranging from large-scale 360-degree panoramic views to microscopic images. Wide-angle cameras, for example, provide an expanded field of view, while fisheye lenses introduce specific distortions for specialized perspectives~\cite{Liang2021AuthTrackEA}. On the other end of the spectrum, microscopic imaging captures minute details beyond what the naked eye can perceive~\cite{hu2023microcam, Dogan2021SensiCutML}. The scale of an image plays a crucial role in determining its application, with panoramic images being essential for XR and mapping, while microscopic images excel in areas such as surface sensing, where detail-oriented analysis is key.

\revise{
\textcolor{Plum}{\textbf{\textit{Dimension-3. Spatial:}}}
The spatial dimension captures depth perception and the three-dimensional (3D) structure of scenes, forming the basis for interpreting and interacting with complex environments. This aligns with Eriksson et al.'s ``space'' concept, emphasizing spatial configurations and dynamic camera spaces in creating immersive systems~\cite{eriksson2007movement}. Depth data from RGB-D cameras or Light Detection and Ranging (LiDAR) provides precise spatial relationships, supporting environmental modeling and navigation~\cite{Wang2024WatchYM, Fleer2012MISOAC}, while computational methods infer depth from 2D images, enabling spatial analysis without dedicated sensors~\cite{Wang2020CAPturARAA, Uimonen2023AGM}.
This dimension also incorporates the dynamic nature of ``camera spaces,'' where user movement and sensor positioning shape spatial perception~\cite{eriksson2007movement}. Integrating spatial data with other modalities, such as IMU or radar, enhances context-aware systems, enabling adaptive real-time interaction in applications like VR, AR, and robotics.
}

\textcolor{Plum}{\textbf{\textit{Dimension-4. Temporal:}}}
The temporal dimension refers to the dynamic aspect of vision, capturing changes in a scene over time. While static images offer a single moment in time, videos record movement and transitions, providing a continuous view of an evolving environment~\cite{Hautasaari2024EmoScribeCA, Shen2019ContextAwareDA}. This dimension enables the analysis of patterns, trends, and behaviors over time, offering essential insights for applications such as autonomous driving, where real-time monitoring of environmental changes is critical. 
\revise{Additionally, event cameras, which capture asynchronous brightness changes at the pixel level, provide a lightweight yet high-resolution representation of motion, enabling efficient tracking of rapid temporal dynamics~\cite{chen2024video2haptics}.
}
By focusing on temporal changes, context-aware systems can track and respond to dynamic elements in the environment, allowing for more adaptive decision-making.

\revise{
\textcolor{Plum}{\textbf{\textit{Dimension-5. Beyond-Human-Vision:}}}
The beyond-human-vision dimension extends perception beyond the visible spectrum, capturing electromagnetic radiation such as infrared, thermal, and ultraviolet waves. Unlike the spatial dimension, which focuses on geometric and depth information within visible light, this dimension broadens the scope by enabling systems to detect otherwise invisible properties of the environment. For instance, thermal cameras can capture heat variations, critical for applications such as object detection and environmental monitoring~\cite{Cho2018DeepTI, Saad2023HotFootFU}. Infrared imaging is frequently used for low-visibility scenarios or detecting surface characteristics, while multispectral imaging leverages light from multiple wavelengths to extract diverse features~\cite{Yeo2017SpeCamSS, Harrison2008LightweightMD}. Additionally, LiDAR technology expands beyond visual light, offering precise 3D mapping and environmental awareness~\cite{Wang2020CAPturARAA, Hwang2020MonoEyeMH}. This dimension is distinct for its ability to capture properties invisible to the naked eye, facilitating novel applications in areas such as healthcare, security, and advanced robotics.
}

\begin{table*}[htbp]
\centering
    \caption{\revise{An overview of visual and other modalities with examples highlighting their roles in enhancing context awareness.}}
    \setlength{\tabcolsep}{6pt}  
    \renewcommand{\arraystretch}{1.8}  
    \resizebox{\linewidth}{!}{
    \begin{tabular}{>{\raggedright\arraybackslash}m{3cm} m{2cm} m{3cm} m{3cm} m{4cm}}  
        \Hline
        \textbf{Examples} & \textbf{Device (Data Source)} & \textbf{Visual Modality} & \textbf{Other Modalities}  & \textbf{Role of Multimodal Data for Enhancing Context Awareness} \\
        \midrule
        Pose-on-the-Go~\cite{Ahuja2021PoseontheGoAU} & smartphone & standard-vision, spatial & motion, haptic & Aiding context awareness in activity tracking, particularly in mobile applications \\
        
        VEmotion~\cite{Bethge2021VEmotionUD} & GPS sensor, smartphone & standard-vision, temporal & audio, text, positional, motion, physiological &  Infering the driver's emotional state indirectly by integrating diverse real-time environmental and vehicular context inputs \\

        Blind Walking Guidance~\cite{Lin2014ACA} & camera, microphone and laser sensor & standard-vision, spatial, beyond-human-vision & audio & Enhancing understanding of object placement and user activity, providing real-time feedback in navigation tasks \\

        MicroCam~\cite{hu2023microcam} & smartphone, server & standard-vision, scale  & motion & Improving surface detection of user actions in location-based services \\

        VirtuWander~\cite{wang2024virtuwander} & VR helmet & standard-vision, spatial & audio, text & Boosting system responsiveness and precision in activity recognition while deepening understanding of user intention \\
        
        EyeMU~\cite{kong2021eyemu} & smartphone & standard-vision, temporal & audio, motion  & Enhancing identity recognition through multimodal interactions such as synchronizing gestures and voice \\
        \Hline
    \end{tabular}
    }
\label{tab:modality_summary}
\end{table*}

\revise{
We provided examples in each category of data modalities and listed them in Table~\ref{tab:visual_example}.
It is important to note that the defined dimensions are not strictly orthogonal; a single image type may exhibit characteristics spanning multiple dimensions. For instance, LiDAR data~\cite{Deng2023ContextAwareFF}, categorized under the spatial dimension, also possesses properties beyond human vision. However, classification is guided by the most prominent feature. Since LiDAR data is primarily used for detecting spatial parameters and often complements camera inputs, we show it in the ``spatial'' category. Similarly, the grayscale image example~\cite{Yang2012MagicFA} in the standard-vision dimension also reflects scale characteristics, as it is captured through a microscope lens.
}

\subsection{Other Modalities} 
Other modalities are categorized from the perspective of sensing devices. These are different types of data that can be collected from various sensors. In this part, we discuss several common sensing modalities that can be combined with camera data to enhance system performance.

\textcolor{Plum}{\textbf{\textit{Modality-1. Audio:}}}
Audio is the modality that is most commonly combined with vision. It serves as both an input and output medium, enhancing the scope of interaction and engagement. As an input, audio can take various forms~\cite{Fleer2012MISOAC, Kianpisheh2024exHAR, xu2024can}. For instance, the human voice can be employed in conjunction with visual information to understand human intention~\cite{Tsai2024GazeNoterCA, Cai2024PANDALensTA, Lee2024GazePointARAC, Wang2024GVOILAGI} and human emotions~\cite{Bethge2021VEmotionUD, Mittal2021MultimodalAC, Lin2024SelfCE, McDuff2019AME}, which is useful in applications like video conferencing~\cite{Hautasaari2024EmoScribeCA}.  

\textcolor{Plum}{\textbf{\textit{Modality-2. Text:}}}
Text is also often combined with the visual modality in VMIs. There are two primary input methods: the first involves using technology, such as software APIs, to convert speech into text, enabling the system to accurately interpret user commands and facilitate more natural, intuitive interactions. This approach is widely applied in areas such as human state understanding~\cite{xu2024can} and education~\cite{Rachabatuni2024ContextawareCU}. The second method involves manual text input, commonly used in content recommendation, creation, and retrieval~\cite{Fan2024ContextCamBC,xu2024can,Liu2023MaTCRMT,Meng2022ValenceAA,Zheng2021StackedMA,Rawat2017ClickSmartAC}.
\revise{Notably, we classify speech-to-text under `text' as it represents processed textual input for system, while human voice remains under `audio' due to its raw acoustic nature, highlighting the complementary roles of these modalities in multimodal integration.
}

\textcolor{Plum}{\textbf{\textit{Modality-3. Motion:}}}
Motion data typically captures parameters such as speed, direction, and acceleration of an object or person. Common sensors include gyroscopes, accelerometers, and Inertial Measurement Units (IMUs), with IMUs being integrated systems that often combine both gyroscopes and accelerometers, along with other sensors. It is often used in conjunction  with camera data for tasks such as motion-tracking~\cite{Hwang2020MonoEyeMH, meyer2022u, Meyer2021ACH, Siddiqi2021AUA}, gesture or pose estimation~\cite{Uimonen2023AGM, Ahuja2021PoseontheGoAU, Wang2024PepperPoseFP}, and surface sensing~\cite{hu2023microcam, Nouredanesh2022EgocentricVD}.

\textcolor{Plum}{\textbf{\textit{Modality-4. Haptic:}}}
Haptic information, the sense of touch, can work in conjunction with the visual modality. As an input, it provides critical contact geometry information, enhancing the system's understanding of user interactions~\cite{Athar2023VisTacTA}. The sensing method of such data can come from user touch~\cite{Ahuja2021PoseontheGoAU} or from machine manipulation based on tactile devices~\cite{wang2021elastic}.

\textcolor{Plum}{\textbf{\textit{Modality-5. Positional:}}}
Radar and GPS are data sensors commonly used to measure position modalities. Radar has many advantages over visual modalities, such as being resistant to occlusion, high test accuracy, and good privacy and security. Therefore, it is often very complementary to visual modalities. The combination of the two is often used in the field of surveying and mapping~\cite{Chou2020EncoderCameraGroundPR} and autonomous driving~\cite{Malawade2022HydraFusionCS}, where multimodality promotes accurate and reliable environmental perception, which is essential for safe and efficient operation. Similarly, autonomous driving is also a typical application of the combination of GPS and vision~\cite{Bethge2021VEmotionUD}.  This combination is also employed for positioning and navigation tasks, enhancing accessibility in barrier-free applications~\cite{Wu2022ContextawareRD}.

\textcolor{Plum}{\textbf{\textit{Modality-6. Physiological:}}}
Physiological data, such as heart rate~\cite{Khan2021PALWA, Zhao2022MediatedAT}, can provide information about a user's emotional state or level of engagement. This category also includes data obtained from EEG, EMG, EDA, and PPG~\cite{Gupta2023SensoryScapeCE, Wen2024AdaptiveVoiceCA}. These different types of physiological data can provide insights into various aspects of a user's physical and emotional state. This data can be combined with camera data to improve user experience or system performance.

\revise{
To sum up, we classified and explored the foundational role of data modalities in defining VMIs in this section, emphasizing the centrality of visual input while briefly addressing other modalities. 
Table~\ref{tab:modality_summary} further illustrates how multimodal data integration enhances context awareness. For example, the Pose-on-the-Go system combines visual modalities with motion and haptic data for activity tracking, demonstrating practical applications in mobile contexts. Similarly, VEmotion integrates visual, audio, and physiological inputs to infer drivers' emotional states, showcasing the synergy of multimodal data in real-time applications. These examples underscore the importance of visual modalities in context-aware systems while highlighting the potential of multimodal integration to address diverse challenges and enable advanced interactions.
}

\section{System Design Foundations} \label{sec: system_design_foundations}
\revise{
With a clear understanding of multimodal data and its contextual definition, researchers need to develop systems capable of processing multimodal data captured from target contexts.
In this section, we present a categorization of system design considerations for the technical implementation of VMIs. 
Specifically, we focus on the following three research questions that guide HCI system research:
(Q1) Where is multimodal data integrated within the system for multimodal processing?
(Q2) How can we process multimodal processing in VMIs?
(Q3) How should the system be evaluated to understand its performance?
Additionally, we faithfully reported the strengths and limitations of approaches at each stage of system design (\autoref{tab: system_arch}, \autoref{tab: data_processing}, and \autoref{tab: eval_strategy}), offering high-level guidance for HCI researchers in designing and implementing VMI systems.
}

\subsection{\reviseSubSectionTitle{Data Integration Stages:}}
\revisesecond{
VMIs are structured around different levels of multimodal data processing, including sensor-level, feature-level, information-level, and hybrid integration systems. These define how the system integrates multimodal data for better performance.
}


\revisesecond{\textcolor{Bittersweet}{\textbf{\textit{\revise{Stage-1. Sensor-level Integration:}}}}} 
Sensor-level systems integrate raw sensing data at the very early stage of the multimodal systems. For example, in MicroCam~\cite{hu2023microcam}, IMU and microscopic visual data are combined for real-time surface detection. PepperPose~\cite{Wang2024PepperPoseFP} fuses IMU data and visual input for full-body pose estimation, enhancing pose perception in dynamic environments and improving adaptability to various spatial directions. VEmotion~\cite{Bethge2021VEmotionUD} integrates vehicle speed, weather, and road type data to predict driver emotions. This early-stage fusion allows systems to leverage multiple perspectives to enhance performance and robustness in complex real-world scenarios.

\revisesecond{
One challenge in integrating multimodal sensor data is synchronizing and calibrating information from disparate sources.
Proper alignment is essential for capturing time-dependent contexts within interactive systems that require real-time functionality.
For example, VEmotion monitors a driver's behavior using multiple sensors to predict their emotional state in real time~\cite{Bethge2021VEmotionUD}.
The system synchronizes data with precise timestamps from different devices, ensuring the effective analysis of temporal patterns.
In MicroCam, IMU and visual data align with each other precisely for effective surface detection by leveraging temporal correlation between motion signals and visual textures to ensure robust and accurate classification~\cite{hu2023microcam}.
}

\begin{table*}
\setlength{\tabcolsep}{4pt}   
\renewcommand{\arraystretch}{1.5}  
\caption{\revise{A summary of key design considerations for the data integration stages in VMIs.}}
\label{tab: system_arch}
\begin{tabular}{>{\raggedright\arraybackslash}m{3cm}ll} 
\Hline
\multicolumn{1}{l}{} & \textbf{Benefits}                                                                                                  & \textbf{Challenges}                                                                                        \\ \hline
\textbf{Sensor-Level}         & \begin{tabular}[t]{@{}l@{}}+ improved raw data quality\\ + early-stage noise reduction\end{tabular}                    & \begin{tabular}[t]{@{}l@{}}- sensor synchronization \\ - latency\end{tabular}                        \\ \hline
\textbf{Feature-Level}        & \begin{tabular}[t]{@{}l@{}}+ richer multimodal representation\\ + mitigates limitations of single modalities\end{tabular}                                                                                    & \begin{tabular}[t]{@{}l@{}}- synchronization of high-dimensional data\\ - handling varying processing speeds\end{tabular} \\ \hline
\textbf{Information-Level}  & \begin{tabular}[t]{@{}l@{}}+ modular processing\\ + better interpretation of complex data\end{tabular} & \begin{tabular}[t]{@{}l@{}}- computational overhead \\ - synchronization issues\end{tabular}         \\ \hline
\textbf{Hybrid}         & \begin{tabular}[t]{@{}l@{}}+ flexibility for diverse tasks\\ + increased robustness\end{tabular}       & \begin{tabular}[t]{@{}l@{}}- system complexity\\ - modular architectures \\ - latency\end{tabular}   \\ \Hline
\end{tabular}
\end{table*}

\revisesecond{\textcolor{Bittersweet}{\textbf{\textit{\revise{Stage-2. Feature-level Integration:}}}}}
Feature-level systems integrate multimodal data in a later-stage of a data processing pipeline -- they fuse the embedding of the raw sensor data from each modality.
In an exemplary pipeline of this type, the system first encodes data from each modality using dedicated encoders.
These modality-specific embeddings are then fused together using a multimodal encoder, facilitating a multimodal representation of a target context.
\revise{For instance, Saad et al.~\cite{Saad2023HotFootFU} combined thermal and visual features from a thermal camera, leveraging the complementary strengths of these modalities to improve robustness in varying lighting conditions, such as low visibility or shadowed environments. Similarly, Rene et al.~\cite{Fender2018VeltAF} integrated depth and RGB data from a camera to construct more accurate 3D representations, which improved spatial perception and object localization. These feature-level combinations enable a more comprehensive understanding compared to using a single modality, particularly by addressing modality-specific limitations through mutual reinforcement.}

\revisesecond{
Feature-level integration requires precise synchronization to ensure compatibility and alignment of data from different modalities, enabling a comprehensive understanding of the target context. Unlike sensor- or decision-level integration, feature-level systems face unique challenges due to the need to combine high-dimensional and abstract data representations from various sensors. These challenges include varying processing speeds across data encoders, the computational demands of handling multimodal features in real-time, and ensuring temporal and spatial alignment of data. For instance, Saad et al. proposed a thermal-RGB system where lighting variations and thermal noise required preprocessing steps such as filtering inconsistent thermal signatures and normalizing RGB inputs~\cite{Saad2023HotFootFU}. Similarly, the Velt system demonstrated the need for precise calibration and alignment when fusing depth and RGB data for accurate 3D scene representation~\cite{Fender2018VeltAF}. Addressing these issues often involves lightweight feature extraction models, dynamic feature prioritization, and robust synchronization mechanisms to maintain performance under diverse conditions. These strategies enable feature-level systems to harmonize disparate data types effectively while managing computational overhead in real-time applications.
}




\textcolor{Bittersweet}{\textbf{\textit{\revise{Stage-3. Information-Level Integration:}}}}
\revise{We defined ``information'' here as human-perceivable semantics like adaptive voice feedback~\cite{Wen2024AdaptiveVoiceCA}, scene descriptions~\cite{de2024llmr}, as opposed to those ``feature'' defined above (e.g., numerical embeddings~\cite{Su2024RASSARRA}).
Information-level systems typically conduct reasoning on such human-perceivable data to facilitate a more explainable processing pipeline.
}
For example, Lim et al.~\cite{Lim2024ExploringCM} integrated environmental factors like brightness and CO2 levels with wide-angle camera data to form a cohesive scene understanding. Similarly, G-VOILA~\cite{Wang2024GVOILAGI} merged gaze-tracking and environmental data, enhancing situational awareness by understanding both the user's interactions and their surroundings.

This architecture allows for modular processing, and it also requires careful synchronization to avoid misalignment between modalities. While offering a more abstract interpretation of complex data, information-level systems can increase computational load, necessitating efficient fusion algorithms and redundancy checks to manage inconsistencies or missing data.

\textcolor{Bittersweet}{\textbf{\textit{\revise{Stage-4. Hybrid Integration:}}}}
Hybrid integration systems combine feature-level, sensor-level, and information-level fusion, enabling flexible integration across multiple processing stages. For example, EmoTour~\cite{Matsuda2018EmoTourME} fuses audio-visual data, physiological signals, and behavioral cues like eye movements to recognize tourist emotions, capturing data at various stages from raw sensor signals to processed behavior.

These systems offer flexibility in data processing by adapting fusion strategies based on system requirements. EmoTour employs both feature-level and information-level fusion, enhancing system robustness by integrating multiple perspectives. Designers must ensure synchronization across modalities while minimizing latency, as well as implement modular architectures to manage the complexity of multi-stage fusion in dynamic environments.

\begin{table*}[h]
\caption{A summary of key design considerations for the multimodal data processing in VMIs.}
\label{tab: data_processing}
\setlength{\tabcolsep}{4pt}   
\renewcommand{\arraystretch}{1.5}  
\begin{tabular}{lll}
\Hline
                               & \textbf{Benefits}                                                                                              & \textbf{Challenges}                                                                                                         \\ \hline
\parbox[t]{3cm}{\textbf{Foundational model APIs}}        & \begin{tabular}[t]{@{}l@{}}+ quick implementation\\ + high accuracy\\ + easy integration\end{tabular} & \begin{tabular}[t]{@{}l@{}}- cloud service dependence\\ - privacy issues\\ - non-real-time processing\end{tabular}  \\ \hline
\parbox[t]{3cm}{\textbf{Developing dedicated ML models}} & \begin{tabular}[t]{@{}l@{}}+ interpretability\\ + on-device processing\end{tabular}                   & \begin{tabular}[t]{@{}l@{}}- high development effort\\ - dataset collection\end{tabular}                           \\ \hline
\parbox[t]{3cm}{\textbf{Heuristic methods}}              & \begin{tabular}[t]{@{}l@{}}+ rapid prototyping\\ + real-time applications\end{tabular}                & \begin{tabular}[t]{@{}l@{}}- low accuracy\\ - low robustness\end{tabular}                                          \\ \Hline
\end{tabular}
\end{table*}

\subsection{Multimodal Data Processing}
In addition to understanding \textit{where} prior work integrate multimodal data, this subsection introduces \textit{how} the existing literature processes the multimodal data to interpret a target context.
Different from existing surveys on multimodal ML algorithms~\cite{zhao2024deep, baltruvsaitis2018multimodal}, we position our scope on provide a taxonomy and strategic guidance for future HCI researchers for their the system implementations.
Different from ML studies that benchmark ML models for higher accuracy, HCI research studies the full system for better user experiences.
Therefore, it is of great importance to guide HCI practitioners on how they can reasonably prototype a design concept with ML or other approaches.
To this end, we categorize approaches used in the existing literature into three classes, shown as follows.

\textcolor{Bittersweet}{\textbf{\textit{Processing-1. Rapid Solution Prototyping via Foundational Model APIs:}}}
\revise{Existing Machine Learning (ML) hubs like Hugging Face and PyTorch Hub contributed to the community with a large number of APIs so that developers can easily reuse for creative applications.
Implementing the system with model APIs allows VMI researchers to rapidly implement data processing pipeline in VMIs without training a dedicated ML model.
For example, MediaPipe provides off-the-shelf solutions for many real-time on-device solutions like pose detection and object tracking so that developers can build real-time apps with minimal engineering effort~\cite{lugaresi2019mediapipe}.
The community has widely utilized such APIs for prototyping new interactive systems, e.g., with facial recognition and gesture detection~\cite{Liu2022VisuallyAwareAC} and with LLM APIs~\cite{zhou2023instructpipe}.
}
Despite these advantages, the use of APIs brings significant concerns. Privacy issues arise when relying on cloud-based services for data processing, especially in sensitive domains. Additionally, foundational models are typically large, making them unsuitable for on-device processing and introducing latency, which reduces responsiveness in real-time applications~\cite{Wang2024WatchYM}.

\textcolor{Bittersweet}{\textbf{\textit{Processing-2. Developing A Dedicated ML Models:}}}
Developing dedicated machine learning models enables tailored multimodal data processing, offering greater control over system performance and privacy.
Traditional ML techniques, such as Support Vector Machines (SVMs) and Random Forests, are widely applied in structured data tasks such as surface sensing and posture estimation~\cite{Yang2012MagicFA, Bethge2021VEmotionUD}.
Thanks to the powerful open-source libraries such as TensorFlow, PyTorch, and scikit-learn~\footnote{Some open-source ML libraries: Tensorflow: \url{https://www.tensorflow.org/.} PyTorch: \url{https://pytorch.org/.} scikit-learn: \url{https://scikit-learn.org/.} }, researchers usually can quickly implement these ML models in a prototypical system.
Additionally, traditional ML models are usually lightweight and explainable, implying that researchers can easily interact with, and debug, a ML-based prototype in real time to understand the new experience.
However, traditional ML methods are less effective when dealing with unstructured data like images and videos.

To process such unstructured multimodal data, researchers usually incorporated a neural network to enhance the perceptual intelligence of a system.
Prior work demonstrated the effectiveness of such approaches in various tasks like scene understanding and gesture recognition, which are critical to spatial-aware and interaction-aware applications~\cite{Schwrer2023Nav2CANAC, Mittal2021MultimodalAC}.
The main reason behind the success of such methods is their ability to automatically encode unstructured data into useful features for task-specific prediction.
However, to develop these ML models, researchers need to acquire large datasets for training, and the collection of datasets in many downstream applications often requires significant effort~\cite{hu2023microcam}.
Additionally, using deep neural networks in an interactive system will inevitably cause computation overhead.
This brings a big challenge to on-device systems which need to perform real-time inference with a power consumption limit~\cite{Zheng2021StackedMA}.

\textcolor{Bittersweet}{\textbf{\textit{Processing-3. Heuristic Methods:}}}
Heuristic methods, relying on rule-based processing, offer a lightweight and fast alternative to machine learning. 
These methods are particularly used for quickly prototyping the design concepts of an interactive paradigm.
This can facilitate rapid conceptual verification by taking humans in the loop in the early-stage system development process.
For example, simple heuristics can involve predefined rules, such as selecting the closest matching depth map based on a straightforward similarity metric, to guide camera localization during bronchoscopic navigation. These rule-based adjustments enable rapid and lightweight prototyping without requiring extensive computational resources, aligning well with early-stage system design needs~\cite{Shen2019ContextAwareDA}.
Once the system design concept is verified, researchers typically choose to invest more development efforts on a concept by, e.g., building a dedicated ML model, and utilizing simple heuristic methods are not suitable for system deployment due to its over-simplified modeling mechanism.

\subsection{Evaluation Strategies} 
Given a context-aware solution by analyzing the multimodal data, how can we understand its performance?
This subsection introduces three kinds of evaluation approaches commonly used in the prior work.

\begin{table*}[]
\setlength{\tabcolsep}{4pt}   
\renewcommand{\arraystretch}{1.5}  
\caption{A summary of key design considerations for the evaluation strategies in VMIs.}
\label{tab: eval_strategy}
\begin{tabular}{lll}
\Hline
                     & \textbf{Benefits}                                                                                                                                                  & \textbf{Limitations}                                                                                                                                    \\ \hline
\parbox[t]{3cm}{\textbf{Demonstration}}        & \begin{tabular}[t]{@{}l@{}}+ rapid evaluation\\ + collecting practical insights in the early stage\end{tabular}                                           & \begin{tabular}[t]{@{}l@{}}- lack of scalability\\ - lack of standard metrics\end{tabular}                                                     \\ \hline
\parbox[t]{3cm}{\textbf{Technical Evaluation}} & \begin{tabular}[t]{@{}l@{}}+ benchmarking with objective metrics\\ + reproducible experiments\\ + understanding technical system performance\end{tabular} & \begin{tabular}[t]{@{}l@{}}- insufficient human factor considerations\\ - limited applicability in the early stage \\ development\end{tabular} \\ \hline
\parbox[t]{3cm}{\textbf{User Evaluation}}      & \begin{tabular}[t]{@{}l@{}}+ understanding realistic user experience\\ + understanding usability\\ + capturing user interaction data\end{tabular}         & \begin{tabular}[t]{@{}l@{}}- high consumption of human effort\\ - difficulty in reproducing the experiment \\ results\end{tabular}             \\ \Hline
\end{tabular}
\end{table*}

\textcolor{Bittersweet}{\textbf{\textit{Evaluation-1. Prototyping  and Demonstration:}}}
Evaluation through demonstration is a technique used to assess how well a system will perform in specific scenarios. The most common approaches identified include prototypes~\cite{Wen2024AdaptiveVoiceCA, Yang2012MagicFA, Cai2024PANDALensTA, Wang2020CAPturARAA, Matsuda2018EmoTourME, Bethge2021VEmotionUD, Chen2023PaperToPlaceTI, Hwang2020MonoEyeMH}, proof-of-concept demonstrations~\cite{Saad2023HotFootFU, Wang2024GVOILAGI, Ahuja2021PoseontheGoAU, Wang2024PepperPoseFP}, and case studies~\cite{Cho2018DeepTI, xu2024can, Liu2023MaTCRMT, Liao2023GPT4EM, Hallyburton2021SecurityAO}. Other approaches include programming by demonstration, where a system learns behaviors by observing human actions and replicating them~\cite{Su2024SonifyARCS}. For instance, in SonifyARCS, the system learns to generate auditory feedback based on user actions without explicit programming. Additionally, showing example applications~\cite{Harrison2008LightweightMD} demonstrates the practical use of a system in real-world scenarios.
Evaluation through demonstration is particularly useful during early-stage development when quick feedback on system performance is required or when the system operates in novel environments that lack established benchmarks. It is ideal for exploratory systems where proof-of-concept or prototype evaluations can reveal the system’s potential in real-world contexts without needing large-scale deployments. This method is also beneficial when evaluating systems designed for highly specific use cases that require situational or contextual understanding rather than standardized testing.

\textcolor{Bittersweet}{\textbf{\textit{Evaluation-2. Technical Evaluation:}}}
Technical evaluation primarily focuses on assessing key performance parameters of the system. The most common approaches include measuring accuracy~\cite{Yang2012MagicFA, Yeo2017SpeCamSS, Cho2018DeepTI, Wang2024WatchYM, Harrison2008LightweightMD, Wang2020CAPturARAA, Fleer2012MISOAC, Hwang2020MonoEyeMH, Dogan2021SensiCutML, hu2023microcam} and time metrics, such as response time or task completion time~\cite{Wen2024AdaptiveVoiceCA, Lee2024GazePointARAC, Chen2023PaperToPlaceTI, Hwang2020MonoEyeMH, Rawat2017ClickSmartAC, Wang2023UbiPhysio}. Additionally, some works evaluate system performance by comparing their results with other systems, for instance, comparing classification algorithms~\cite{hu2023microcam, Wang2024WatchYM, Meyer2021ACH}. Many studies also employ ablation studies to gain deeper insights into the interface by measuring how each component contributes to the system’s overall performance~\cite{Wang2024WatchYM, Wang2024GVOILAGI, xu2024can, Suo2020Neural3DLN}.
A technical evaluation is particularly suited for later stages of development when the system is relatively stable and requires precise measurements of its effectiveness and efficiency. It is essential when a system is intended to replace or outperform existing solutions, as comparative evaluations and ablation studies offer insights into the system’s strengths and potential weaknesses. Moreover, technical evaluations are ideal when fine-tuning system performance is necessary, as they allow for detailed analysis of accuracy, speed, and individual system components under controlled conditions.

\textcolor{Bittersweet}{\textbf{\textit{Evaluation-3. User Evaluation:}}}
User evaluation refers to measuring the effectiveness of a system through user studies. 
To quantitatively understand the system performance perceived by users, the community has introduced various Likert-scale metrics targeting different evaluation scenarios~\cite{Fan2024ContextCamBC, Su2024SonifyARCS, Cai2024PANDALensTA, Wang2020CAPturARAA, Chen2023PaperToPlaceTI, xu2024can, Wang2024GVOILAGI}, such as  SUS~\cite{Hautasaari2024EmoScribeCA, Lee2024GazePointARAC, Wang2020CAPturARAA, Chen2023PaperToPlaceTI} and NASA TLX~\cite{Hautasaari2024EmoScribeCA, Chen2023PaperToPlaceTI, Tsai2024GazeNoterCA, Liang2021AuthTrackEA} .
Understanding users' qualitative comments also plays a critical role in the evaluation.
Common approaches include conducting interviews~\cite{Lim2024ExploringCM, Cai2024PANDALensTA, Jamonnak2021GeoContextAS} and gathering user feedback~\cite{Su2024SonifyARCS, Wang2020CAPturARAA} through self-reported experiences~\cite{Wang2024PepperPoseFP}, think-aloud studies~\cite{zimmerer2022case}, and diary studies~\cite{Lee2024GazePointARAC}. 

It is important to note that it is a common practice to combine user evaluation with demonstration~\cite{xu2024can, Jamonnak2021GeoContextAS, Liao2023GPT4EM} or technical evaluations~\cite{Wen2024AdaptiveVoiceCA, Harrison2008LightweightMD, Rawat2017ClickSmartAC, Liang2021AuthTrackEA}.
The critical factor in study design is the evaluation's objective -- specifically, the research questions researchers aim to explore.
For example, the System Usability Scale (SUS)~\cite{brooke2013sus} and NASA TLX~\cite{hart2006nasa} are widely used examples of Likert scale-based questionnaires, focusing on measuring usability and perceived workload, respectively~\cite{joshi2015likert}.
Interviews and think-aloud studies are effective when deeper, qualitative insights into user behavior and preferences are needed.
Additionally, think-aloud or task-based studies are well-suited for systems requiring real-time interaction analysis, while surveys or interviews can capture users' overall experiences after interacting with the system. 
A combination of these evaluation techniques usually provides a more comprehensive evaluation but it causes extra workload for researchers.

\section{Application Domains}  \label{sec: application_domains}
\begin{figure*}
    \centering
    \includegraphics[width=0.6\linewidth]{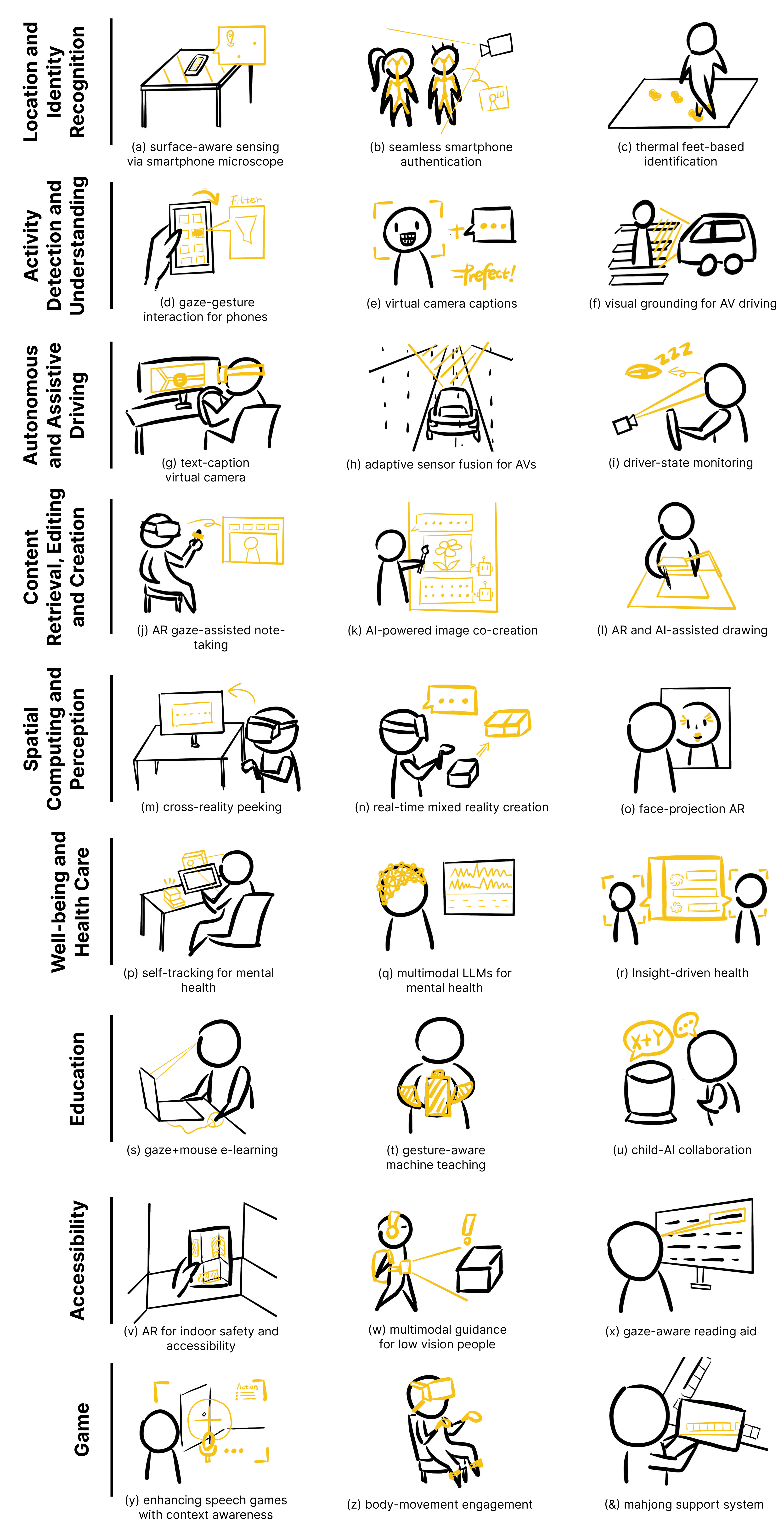}
    \caption{\revise{Examples of application domains for VMIs (illustrative references (a):~\cite{hu2023microcam}, (b):~\cite{Liang2021AuthTrackEA}, (c):~\cite{Saad2023HotFootFU}, (d):~\cite{kong2021eyemu}, (e):~\cite{Hautasaari2024EmoScribeCA}, (f):~\cite{Liao2023GPT4EM}, (g):~\cite{Wen2024AdaptiveVoiceCA}, (h):~\cite{Malawade2022HydraFusionCS}, (i):~\cite{Doudou2019DriverDM}, (j):~\cite{Tsai2024GazeNoterCA}, (k):~\cite{Fan2024ContextCamBC}, (l):~\cite{hoang2024artvista}, (m):~\cite{Wentzel2024SwitchSpaceUC}, (n):~\cite{de2024llmr}, (o):~\cite{bokaris2019light}, (p):~\cite{Lim2024ExploringCM}, (q):~\cite{hu2024exploring}, (r):~\cite{Englhardt2023FromCT}, (s):~\cite{Zhu2023IntegratingGA}, (t):~\cite{zhou2022gesture}, (u):~\cite{Zhang2024MathemythsLL}, (v):~\cite{Su2024RASSARRA}, (w):~\cite{Wang2024GazePromptEL}, (x):~\cite{Wang2024GazePromptEL}, (y):~\cite{zargham2024know}, (z):~\cite{somarathna2023exploring}, (\&):~\cite{Suzuki2019AnOM}).}}
    \label{fig:application}
\end{figure*}

\revise{
We have identified nine key application areas for context awareness in VMIs, selected for their relevance in demonstrating the practical applications and unique capabilities of VMIs. These domains illustrate how VMIs integrate visual and multimodal data to address specific context-aware challenges, emphasizing strengths such as precision, adaptability, and real-time responsiveness. Covering a range of scenarios from location sensing to healthcare and gaming, these areas reflect the practical value of VMIs in supporting context-aware interactions. The selected domains also provide insight into how VMIs contribute to advancing HCI by addressing current challenges and enabling more effective system designs. Detailed examples and applications are shown in Figure~\ref{fig:application}.
}

\textcolor{Dandelion}{\textbf{\textit{Domain-1. Location and Identity Recognition:}}}
Location and identity recognition are key application areas for VMIs. For location recognition, various sensor-based systems have been developed, especially wearable devices like handheld systems (e.g., MicroCam~\cite{hu2023microcam}, SpeCam~\cite{Yeo2017SpeCamSS}, SpectroPhone~\cite{schrapel2021spectrophone}) and leg-mounted devices (e.g., HotFoot~\cite{Saad2023HotFootFU}, RadarFoot~\cite{elvitigala2023radarfoot}). These systems use visual and non-visual modalities to enhance accuracy and environmental awareness. For identity recognition, systems such as Auth + Track~\cite{Liang2021AuthTrackEA} leverage mobile phones and embedded cameras, effectively using fragmented "downtime" between interactions to ensure seamless experiences with minimal user burden. These applications emphasize high recognition accuracy and low latency to maintain system reliability and engagement.

\textcolor{Dandelion}{\textbf{\textit{Domain-2. Activity Detection and Understanding:}}}
VMIs improve activity detection through wearable technologies that recognize actions like gestures and postures, enabling intuitive interaction~\cite{Wang2024PepperPoseFP,zhou2022gesture}. Vision-based gesture recognition is critical, with cameras detecting gestures in mixed reality and interactive learning~\cite{zhou2022gesture}. IMUs (e.g., accelerometers) and sensors further support gesture detection and activity recognition~\cite{kong2021eyemu,fukasawa2022spatial}. Gaze tracking using eye-tracking or neural networks~\cite{Wu2022ContextawareRD, Zhu2023IntegratingGA} monitors attention and cognitive load, enabling adaptive interfaces and intuitive interactions~\cite{Lee2024GazePointARAC, Wen2024AdaptiveVoiceCA,kubo2017syncro,su2021gaze+}. VMIs also detect movements~\cite{Wu2022ContextawareRD}, speech~\cite{Hautasaari2024EmoScribeCA}, and driving behaviors~\cite{Liao2023GPT4EM} across diverse contexts like healthcare and video conferencing~\cite{Mittal2021MultimodalAC, Siddiqi2021AUA}. By integrating multimodal data, VMIs enhance detection accuracy and expand interaction possibilities, making them essential for context-aware systems.

\textcolor{Dandelion}{\textbf{\textit{Domain-3. Autonomous and Assistive Driving:}}}
Driver state detection is vital for safety in autonomous and assistive driving. Many systems assess the driver’s state using behaviors like steering, pedal usage, and vehicle speed~\cite{Hs2010AssessingTF}. Recent systems integrate cameras to monitor gaze, facial expressions, and head movements, improving accuracy~\cite{Koch2023LeveragingDV, Doudou2019DriverDM, Thandlam2024NextGenVS}. Combining these visual cues with in-car sensors detecting conditions like fatigue or intoxication enhances state classification. For example, sensors identifying slurred speech, slow reactions, or alcohol odor improve driver monitoring. External factors like road conditions, weather, and traffic further refine assessments~\cite{Malawade2022HydraFusionCS}. Addressing stress due to challenging conditions rather than impairment enhances safety. Beyond detection, adaptive interventions like visual alerts~\cite{Kunze2019AutomationTI} and voice assistants~\cite{Wen2024AdaptiveVoiceCA, Huang2024ChatbotAF} improve alertness, while haptic feedback (e.g., steering wheel vibrations) enhances response effectiveness. These innovations underscore VMIs’ role in improving driving safety through context-aware interventions.

\textcolor{Dandelion}{\textbf{\textit{Domain-4. Content Retrieval, Editing and Creation:}}}
Multimodal interactions are increasingly essential in VMIs, enabling systems to interpret complex contexts effectively. Integrating multimodal LLMs allows richer interactions by processing inputs from visual, audio, and textual data. For example, facial expressions and gaze~\cite{Tsai2024GazeNoterCA} enhance user intent interpretation, critical for creative applications where text-based inputs limit design exploration. Combining multimodal data improves context understanding, supporting adaptive interaction in domains like education and entertainment~\cite{Fan2024ContextCamBC}. GenAI further reduces creation costs by generating initial content via Diffusion models, enabling users to refine outputs~\cite{hoang2024artvista}. By advancing context awareness, multimodal VMIs facilitate intuitive, dynamic, and user-centered interactions, paving the way for more flexible human-computer collaboration.

\textcolor{Dandelion}{\textbf{\textit{Domain-5. Spatial Computing and Perception:}}}
VMIs play a critical role in enhancing context awareness in XR environments. Advancements in tracking technologies have introduced intuitive interactions like peeking~\cite{Wentzel2024SwitchSpaceUC}, body-around~\cite{ferdous2019s,bokaris2019light}, object-centric~\cite{liao2024realityeffects}, bare-hand~\cite{Kim2024QuadStretcherAF}, audio~\cite{molina2023storytelling}, and text-based interactions~\cite{de2024llmr,shoa2023sushi,chen2024supporting,yin2024text2vrscene}. For example, modern LLMs revolutionize text-based workflows by enabling efficient and intuitive user interactions. However, challenges like noisy real-time tracking hinder precise virtual alignment~\cite{li2024predicting}. Addressing these alongside improving display and tracking technologies could expand XR’s precision applications, such as surgical assistance. Collaborative XR environments are advancing but require overcoming latency and network distortion to ensure seamless teamwork~\cite{Jin2023CollaborativeOL, Sameri2024CollaborativeCI}. VMIs’ continued development will enhance spatial perception, enabling accurate, context-aware interactions across industries.

\textcolor{Dandelion}{\textbf{\textit{Domain-6. Well-being and Health Care:}}}
The integration of VMIs and LLMs holds great promise for healthcare by enhancing context awareness through multimodal data analysis. While LLMs are widely used to infer mental~\cite{Xu2023MentalLLM} and physical~\cite{Jin2024HealthLLMPR, Englhardt2023FromCT} health from text-based data, incorporating visual information like facial expressions and video-based emotion tracking can provide more holistic patient assessments~\cite{hu2024exploring}. Combining visual and physiological signals (e.g., EEG, EPG) offers critical insights, especially in mental health care, where body language and expressions reveal psychological states~\cite{Lim2024ExploringCM}. Integrating VMIs into LLM-driven healthcare systems enables more effective diagnosis, monitoring, and personalized treatment by leveraging multimodal inputs, unlocking new healthcare innovations.

\textcolor{Dandelion}{\textbf{\textit{Domain-7. Education:}}}
With advances in AI, VMIs have become essential for educational applications by integrating visual and non-visual modalities like text, sound, and sensors to create personalized learning experiences. For instance, VMIs enable intuitive interactions by interpreting gestures, voice, and contextual cues~\cite{Zhang2024MathemythsLL, zhou2022gesture,Zhu2023IntegratingGA}. Gesture-aware systems like LookHere use real-time visual feedback to enhance machine teaching~\cite{zhou2022gesture}, enabling learners to annotate objects via natural gestures. Multimodal LLMs further enhance these interfaces by interpreting complex data and generating contextually relevant information, improving engagement in remote learning and multimedia interactions~\cite{Rachabatuni2024ContextawareCU}. Combining gaze and mouse data improves engagement monitoring in e-learning, outperforming single-modality methods~\cite{Zhu2023IntegratingGA}, demonstrating VMIs’ potential to personalize and optimize educational technologies.

\textcolor{Dandelion}{\textbf{\textit{Domain-8. Accessibility:}}}
VMIs have significantly advanced accessibility by combining visual and other sensory modalities to facilitate seamless interaction and navigation. For instance, integrating visual data with auditory or haptic feedback enhances navigation for visually impaired users in digital and physical environments~\cite{Wang2024GazePromptEL, Su2024RASSARRA}. However, challenges remain in ensuring multimodal systems’ reliability in real-world scenarios where sensor data may be disrupted. Addressing issues like noise, interference, and seamless integration of multiple modalities requires robust algorithms and efficient designs~\cite{Lin2014ACA}. Solving these challenges is crucial for developing accessible systems that ensure reliable, user-friendly interactions across diverse settings.

\textcolor{Dandelion}{\textbf{\textit{Domain-9. Game:}}}
 In gaming, VMIs enhance context awareness by integrating visual data with modalities like sound~\cite{zargham2024know}, motion, and physiological signals~\cite{somarathna2023exploring}, creating immersive environments. In VR games, combining sensor-tracked body movements with visual data assesses user engagement and immersion~\cite{zargham2024know,somarathna2023exploring}. AR applications like Mahjong use context-aware image recognition to provide real-time feedback and improve gameplay efficiency~\cite{Suzuki2019AnOM}. Integrating modalities like IMUs or haptic feedback further enhances interaction by contextualizing users' physical and virtual surroundings, enabling adaptive gaming experiences that respond to real-time contexts.

\section{Design Considerations and Key Challenges}
\label{sec: design_challenges}

\revise{
Following the steps outlined above, a general context-aware system can be constructed. Revisiting the framework is necessary to ensure critical design considerations are addressed. These considerations are categorized into three aspects: user-centric considerations (challenges 1, 2, 3, and 4), data management and processing (challenges 5, 6, and 7), and system integration and resource optimization (challenges 8-12). This classification provides a structured approach to identify key aspects and guide future refinements.
}

\textcolor{ForestGreen}{\textbf{\textit{Challenge-1. Privacy and Security-aware Systems:}}} 
\revise{
Privacy and security are critical considerations in VMI systems, as evidenced by 46 related studies, highlighting the potential sensitivity of visual data. Such data often contains rich, personal information, necessitating robust measures to protect user privacy. For example, PAL~\cite{Khan2021PALWA} used on-device deep learning for privacy-preserving, low-shot context detection, while Wang et al.~\cite{Wang2024WatchYM} developed a silent-speech interaction system that avoids visual data collection by relying on depth sensing. Despite these advancements, key challenges persist, particularly in minimizing the amount of data collected and ensuring the protection of sensitive visual information. 
Strategies to address these challenges include dynamically adjusting the level of data detail~\cite{padilla2015visual}, capturing only essential silhouette information~\cite{li2020towards}, and using obfuscation techniques such as perturbations~\cite{ye2022visual} and style transfer~\cite{xu2024examining}. However, the ongoing arms race between privacy-preserving measures and increasingly sophisticated extraction attacks necessitates continuous development of more robust defense strategies.
}

\textcolor{ForestGreen}{\textbf{\textit{Challenge-2. User Variability:}}} 
\revise{
User variability is discussed in 51 studies, reflecting its significance in the design of VMI systems. Diverse user characteristics, such as facial structure, head shape, or personal preferences, can impact system performance and accuracy. For instance, RASSAR~\cite{Su2024RASSARRA} focused on incorporating varying accessibility requirements when designing systems, emphasizing the differences in users' indoor environments and the need for customized workflows to evaluate safety risks. Similarly, Ahuja et al.~\cite{Ahuja2021PoseontheGoAU} addressed personal variance in full-body pose estimation by employing extensive sensor fusion, utilizing both front and rear cameras on smartphones. 
Despite these advancements, user variability remains a significant challenge for VMI systems. Variations in facial structure, head shape, and other physical features can significantly affect sensor accuracy~\cite{ban2020ear, fu2020systematic, lee2018anthropometric}. Moreover, environmental factors, including lighting conditions and occlusions such as glasses or piercings, further complicate system performance~\cite{bedri2017earbit, laput2016sweepsense, min2019early, huang2016wearable}.
}

\revise{
\textcolor{ForestGreen}{\textbf{\textit{Challenge-3. Ethics:}}}
In our survey, 27 studies have specifically examined challenges related to LLMs~\cite{Wang2024PepperPoseFP} and embodied AIs~\cite{de2024llmr}. These works focused on the relationship between humans and AIs, addressing concerns such as data privacy, the protection of human rights during experimentation, and the potential consequences for future AI development. For instance, De et al.~\cite{de2024llmr} explored fears around job displacement for developers and creators but highlighted that their framework facilitates more effective human-AI collaboration by ensuring human involvement in the system. Despite these efforts, key ethical challenges remain, including the need to mitigate algorithmic bias, ensure fairness in AI-driven decision-making, and protect sensitive data. Furthermore, open challenges persist in addressing the societal implications of AI systems, such as ensuring transparency, accountability, and the responsible use of technology, particularly in contexts involving vulnerable populations.
}

\revise{
\textcolor{ForestGreen}{\textbf{\textit{Challenge-4. Cognitive Load and User Engagement:}}}
Cognitive load and user engagement are closely interconnected, mentioned in 58 work, often exhibiting a negative correlation. These factors are crucial considerations in the design of action-intensive systems, particularly within immersive environments like VR. One of the primary challenges highlighted is the difficulty in quantifying and managing cognitive load, which directly impacts user engagement. Users often face barriers to sustained interaction due to the overwhelming complexity of VR environments. The integration of multimodal data and enhanced context awareness present valuable opportunities to address this challenge.
For example, Somarathna et al.~\cite{somarathna2023exploring} introduced body movements as a novel indicator of user engagement in VR gaming, offering an alternative to traditional methods. Wen et al.~\cite{Wen2024AdaptiveVoiceCA} developed a cognitively adaptive voice interface that adjusts information delivery based on varying levels of urgency and cognitive demand, aiming to optimize the balance between system responsiveness and user cognitive load. However, despite these advancements, key challenges persist. These include the need for more precise methods to measure cognitive load in real-time and across diverse users, as well as the challenge of maintaining user engagement without leading to mental fatigue. Additionally, the development of systems that can dynamically adapt to individual cognitive states throughout an interaction remains an unresolved issue.
}

\textcolor{ForestGreen}{\textbf{\textit{Challenge-5. Automated Sensor Configuration:}}}
\revise{
Sensing is fundamental to capturing vision-based multimodal data, reflected in 32 studies. }
The rapid growth of the IoT, which connects billions of sensors, has made manual sensor configuration impractical~\cite{perera2013dynamic}. To address this, several works have focused on automating or semi-automating sensor connections to applications. For example, Kong et al.~\cite{kong2021eyemu} explored the automatic integration of gaze-tracking sensors and IMUs to recognize gestures, while Ahuja et al.\cite{Ahuja2021PoseontheGoAU} used dynamic deployment of sensors and cameras in conjunction with inverse kinematics algorithms to estimate full-body poses. These efforts laid the foundation for automating sensor configuration, particularly in systems designed for human recognition. Despite these advances, several key challenges remain. These include the need for more robust methods for automatic sensor discovery, seamless integration of diverse sensor types, and ensuring real-time data synchronization across large-scale sensor networks.

\textcolor{ForestGreen}{\textbf{\textit{Challenge-6. Context Discovery:}}}
\revise{This aspect was mentioned by 48 studies, focusing on how to automatically interpret and annotate sensor data within diverse application domains.} 
As sensor data is generated, it must be contextualized to make it meaningful and actionable. Several approaches have been proposed to automate this process. For instance, Zargham et al.~\cite{zargham2024know} explored context-aware speech recognition, where environmental and action-based context from the game enhanced the accuracy of speech recognition in interactive gaming. Similarly, Su et al.~\cite{Su2024SonifyARCS} utilized event context during AR interactions, applying LLMs and audio models for context-based sound acquisition and sonification. These efforts highlight the potential of integrating context awareness into sensor systems, but challenges remain. Specifically, there is the difficulty of automating context discovery in highly heterogeneous environments, where sensor data may vary significantly across different domains. While advances in semantic technologies and linked data~\cite{wang2004ontology, compton2009survey, spanos2012sensorstream, heath2011synthesis, le2009linked} offer promising avenues for future development, key open challenges include improving the accuracy and scalability of context annotation across diverse applications and enabling real-time context awareness in dynamic environments.

\textcolor{ForestGreen}{\textbf{\textit{Challenge-7. Semantic Multimodal Data Integration:}}}
\revise{A total of 54 papers have discussed this related aspect.}
\revise{Multimodal systems often require the semantic alignment of diverse sensor modalities, necessitating the development} of advanced semantic modeling frameworks~\cite{wang2004ontology}. For example, Wang et al.~\cite{Wang2023UbiPhysio} developed a companion bot with a visual interface that semantically integrates sensor data to provide enhanced feedback for rehabilitation, while Xu et al. used multimodal fusion of visual and audio signals through LLMs to create evolving user profiles in conversational agents~\cite{xu2024can}. Despite these efforts, \revise{semantic integration of multiple modalities often remained a case-specific manner, which scalability of the fusion algorithms across diverse sensor types and contexts, or the development of adaptive fusion algorithms worth of future exploration.}

\begin{figure*}
    \centering
    \includegraphics[width=\linewidth]{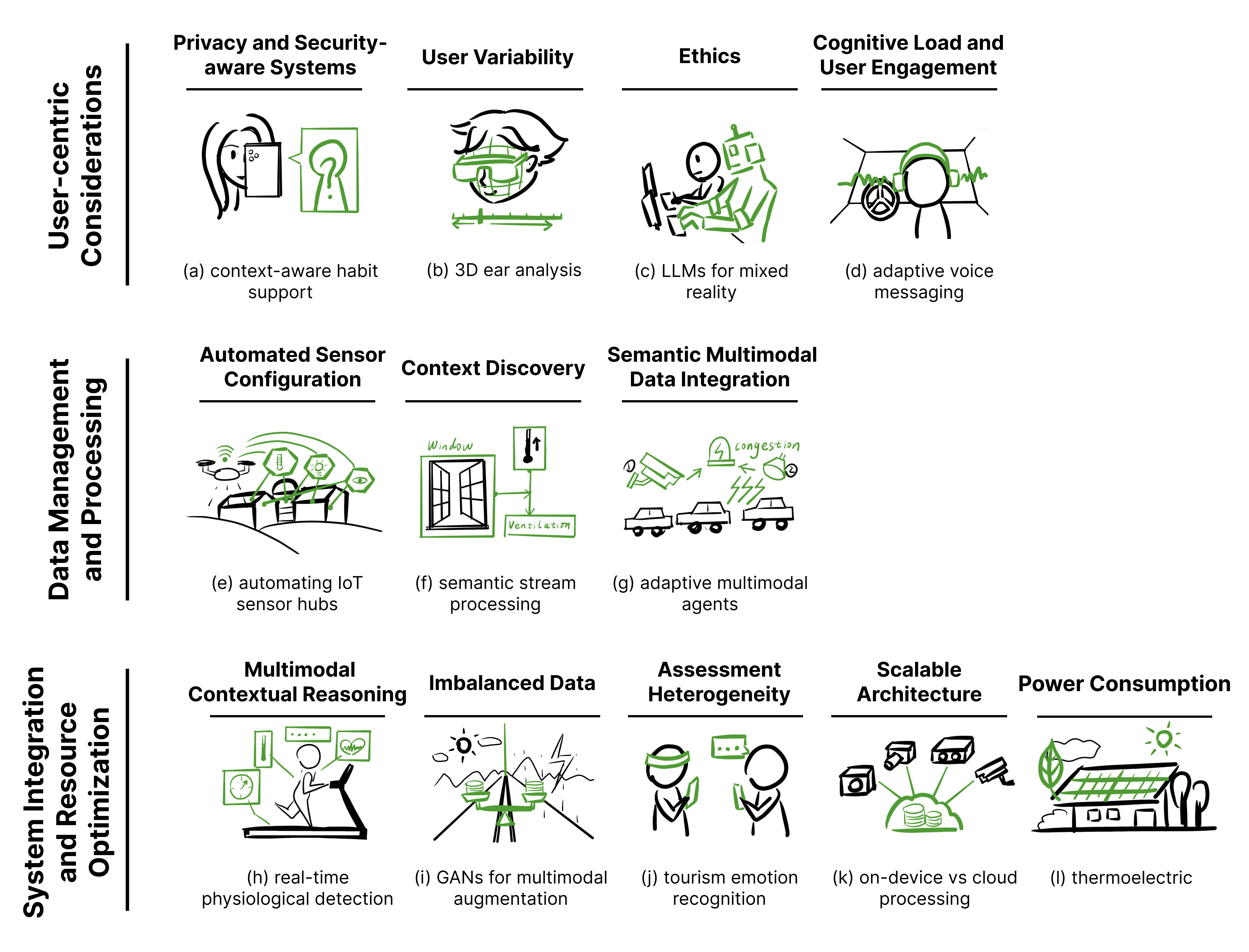}
    \caption{\revise{Examples of design considerations and key challenges for VMIs (illustrative references (a):~\cite{Khan2021PALWA}, (b):~\cite{lee2018anthropometric}, (c):~\cite{de2024llmr}, (d):~\cite{Wen2024AdaptiveVoiceCA}, (e):~\cite{perera2013dynamic}, (f):~\cite{spanos2012sensorstream}, (g):~\cite{Liao2023GPT4EM}, (h):~\cite{Koch2023LeveragingDV}, (i):~\cite{pahde2021multimodal}, (j):~\cite{Matsuda2018EmoTourME}, (k):~\cite{McDuff2019AME}, (l):~\cite{bouchard2020ear}.}}
    \label{fig:}
\end{figure*}

\textcolor{ForestGreen}{\textbf{\textit{Challenge-8. Multimodal Contextual Reasoning:}}}
\revise{The reasoning-related processing was involved in 73 tasks.}
\revise{In VMI systems, contextual reasoning enables the system to interpret complex, dynamic interactions, enhancing its ability to understand and respond to evolving situations. However, reasoning about the relationships is inherently context and task-dependent.} Koch et al.~\cite{Koch2023LeveragingDV} developed a system for inferring blood alcohol concentration in real-time based on gaze and head movement data, while Fan et al.~\cite{Fan2024ContextCamBC} integrated contextual reasoning into a human-AI co-creation system for generating artistic images. Although there were early attempts and early advancements, significant challenges remain particularly \revise{in achieving real-time reasoning with high computational efficiency, which however is often essential for VMI systems' deployment. Additionally, most algorithms faced the challenges of improving their adaptability on evolving new contexts and accurately and proactively reason about users' intentions.}

\textcolor{ForestGreen}{\textbf{\textit{Challenge-9. Imbalanced Data:}}}
\revise{27 papers have explored the issue of class imbalance in VMIs,} 
\revise{Data are often imbalanced in real-world scenarios, especially for those detection-based VMI systems, resulting in biased performance~\cite{he2009learning} or even failure to deploy. Researchers proposed several sampling and training methods, specifically targeted at maintaining VMI systems' accuracy.} For instance, random oversampling and undersampling have been applied to balance classes, though oversampling can lead to overfitting, and undersampling may discard valuable data~\cite{lee2019machine}. More advanced techniques, such as SMOTE~\cite{chawla2002smote}, have been used to generate synthetic examples by interpolating between instances, though careful feature normalization is required to avoid introducing noise. Additionally, Generative Adversarial Networks have been used to augment multimodal datasets, such as adding missing text data paired with visual inputs~\cite{pahde2021multimodal}. However, as VMI systems often operate in dynamic environments with evolving data distributions, further research is needed to explore how these techniques can adapt to such changing contexts over time.

\textcolor{ForestGreen}{\textbf{\textit{Challenge-10. Assessment Heterogeneity:}}}
\revise{A total of 11 papers have discussed the challenge of assessment heterogeneity in VMI systems, particularly in comparing the performance of different sensing modalities.}
\revise{Inconsistent evaluation protocols and metrics across different sensing modalities often hinder progress in this area, as they prevent meaningful comparisons between studies and affect the ability to benchmark VMI systems accurately~\cite{ne2021hearables}. This variability directly impacts VMI systems, where multiple modalities (e.g., vision, speech, and motion) need to be integrated and evaluated cohesively. While studies like those by Matsuda et al.\cite{Matsuda2018EmoTourME} and Sun et al.~\cite{Sun2023TemporallyCS} have made strides toward aligning emotion recognition performance metrics across modalities and ensuring consistency in segmentation, the absence of unified evaluation standards remains a fundamental barrier. Establishing clear, standardized protocols for VMI system evaluation is essential to facilitate accurate assessments, foster meaningful comparisons, and drive further innovations in the field. }

\textcolor{ForestGreen}{\textbf{\textit{Challenge-11. Scalable Architecture:}}}
\revise{Scalable architecture is essential for evaluating a system, especially in the context of VMIs where different modalities greatly increased the complexity of the system. This aspect is highlighted by 63 papers, where many researchers specifically examined the trade-off between on-device and cloud-based processing~\cite{McDuff2019AME}, or explored the scability in specific fields such as autonomous driving or tasks such as 3D object detection~\cite{Deng2023ContextAwareFF}. With the increasing integration of VMIs within IoT ecosystem, scalable}, distributed architectures are crucial for managing heterogeneous sensor modalities, the high computational demands and real-time data processing requires. Cloud-edge hybrid models, which balance resource allocation across local and cloud systems, hold promise for addressing these needs. However, open challenges remain in optimizing these architectures for large-scale deployments, ensuring real-time data handling, and maintaining adaptability as IoT devices evolve.

\textcolor{ForestGreen}{\textbf{\textit{Challenge-12. Power Consumption:}}}
\revise{28 papers have addressed power consumption issues in context-aware systems,} which are critical due to the high computational demands of real-time visual processing. 
Continuous tasks like image recognition or large-scale data analysis are \revise{common in VMIs, however require significant power,} particularly for mobile or wearable systems constrained by battery life. Techniques such as dynamic frame rate adjustments and energy-efficient image sensors \revise{accordingly targeted} at reducing power drain during data collection. For example, Chen et al.~\cite{chen2024video2haptics} utilized event-based visual sensing, inspired by biological systems, to achieve low data rates and reduced power consumption. Zhang et al.~\cite{zhang2024earsavas} proposed a subject-aware vocal activity sensing method that reduces power usage by avoiding unnecessary system wake-ups. Additionally, energy harvesting technologies, such as thermoelectric or motion-based generators~\cite{bouchard2020ear, delnavaz2013piezo}, show potential for extending battery life in real-world applications. \revise{Despite these progress,} ongoing challenges remain in optimizing energy efficiency without compromising performance, especially in systems requiring continuous operation.

\section{Findings and Discussions}\label{sec:findings_discussions}

\begin{figure*}
    \centering
    \includegraphics[width=1.2\linewidth, angle=-90]{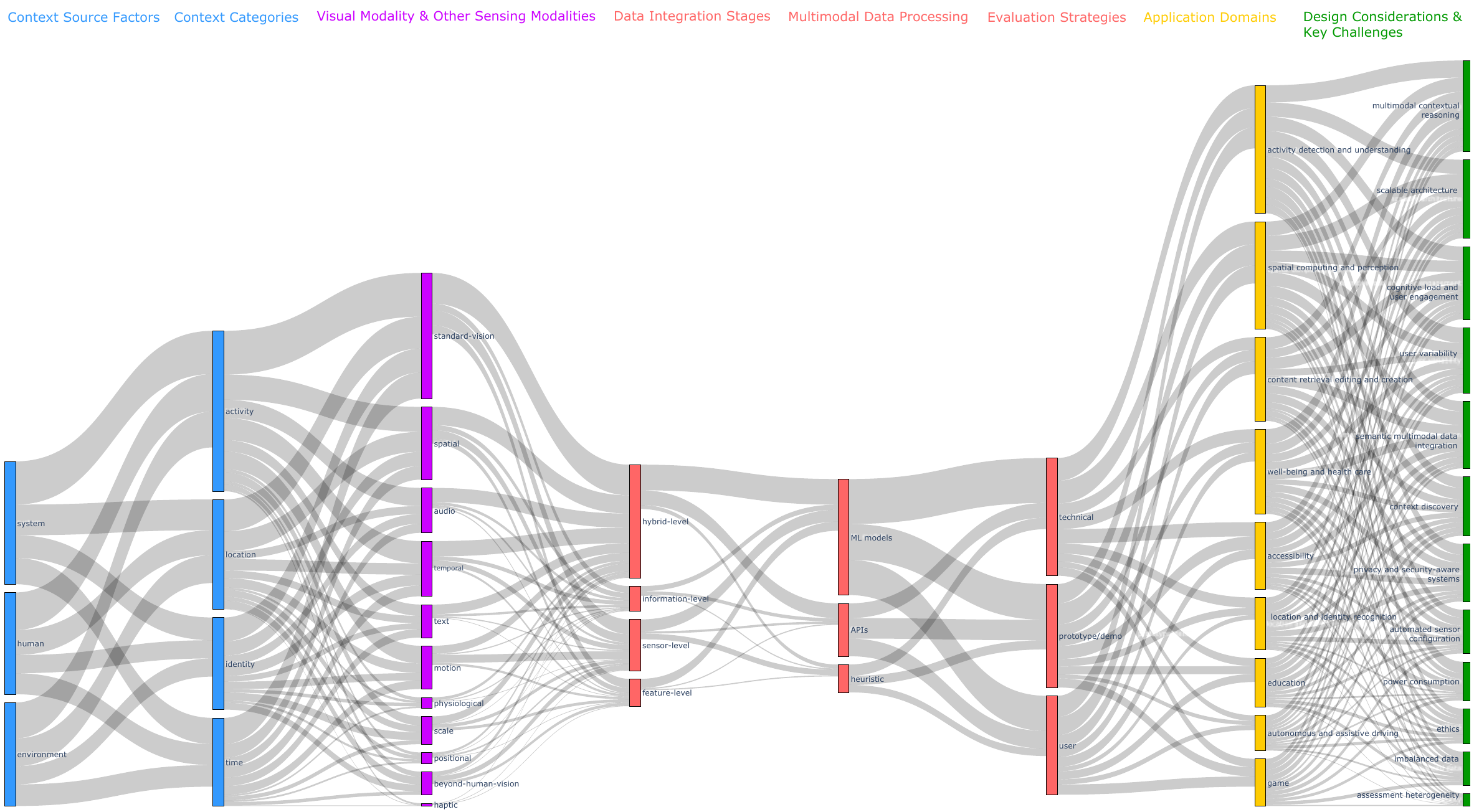}
    \caption{\revisesecond{A Sankey diagram summarizing the overall literature counts across critical dimensions of our taxonomy. From top to bottom (from left to right), the columns represent: context source factors, context categories, visual and other sensing modalities, data integration stages, multimodal data processing, evaluation strategies, application domains, design considerations, and key challenges.}}
    \label{fig:sankey}
\end{figure*}

We conducted a literature survey to statistically analyze the application of VMIs in context-aware systems, presenting the results in a Sankey diagram (Figure~\ref{fig:sankey}). An interactive version is also available in HTML format\footnote{Source files and an interactive Sankey diagram are accessible at: \url{https://drive.google.com/drive/folders/18dNSB9JuftudTCZss_yeJsvwflYYSJvN?usp=sharing}.}. This visualization facilitates detailed exploration and querying of the data. It intuitively illustrates the flow of contextual information across critical dimensions, offering insights into relationships among taxonomy elements.

\subsection{Node Analysis}
The Sankey diagram highlights key nodes. For instance, ``System Factors'' in Context Source Factors is referenced in 106 studies, representing 97\% of the surveyed literature. This underscores its essential role in achieving resource-efficient and scalable VMIs. For example, MicroCam~\cite{hu2023microcam} integrates visual and IMU data to enhance context awareness in resource-constrained settings. Similarly, the ``Environmental Factors'' node, cited in 92 references, emphasizes the need for adaptive algorithms to address environmental variability across dimensions like ``Activity,'' ``Location,'' ``Identity,'' and ``Time.'' For instance, systems designed to dynamically respond to environmental changes, such as lighting or noise conditions~\cite{bokaris2019light}, can improve alignment between virtual and physical environments. Practitioners should prioritize adaptive capabilities to enhance the robustness of multimodal interactions under real-world conditions.

In terms of application domains, we identified four prominent areas: ``Activity Detection and Understanding,'' ``Content Retrieval, Editing, and Creation,'' ``Spatial Computing and Perception,'' and ``Well-being and Healthcare.'' Activity detection, exemplified by EyeMU~\cite{kong2021eyemu}, leverages visual and motion data for real-time user action interpretation, addressing variability across contexts. Content creation applications utilize multimodal inputs like gaze and gestures to enhance user intent interpretation and streamline workflows. In spatial computing, systems like GazePointAR~\cite{Lee2024GazePointARAC} align virtual and physical elements, though challenges like latency and noise remain. Healthcare applications, such as VEmotion~\cite{Bethge2021VEmotionUD}, integrate visual and physiological signals for holistic patient monitoring. These examples highlight the need for adaptive algorithms and robust multimodal fusion to ensure scalability and effectiveness in diverse, real-world applications.

\subsection{Pathway Analysis}
\revise{
Figure~\ref{fig:sankey} visualizes the flow of information across critical dimensions, revealing key pathways that emphasize the role of visual modalities in context-aware systems. These pathways highlight two perspectives: the standalone contributions of visual modalities to activity detection and their integration with location-based data for spatial computing.
}

\revise{
One critical pathway, ``Activity-Standard-Vision-Sensor-Level-ML Models-Activity Detection and Understanding,'' underscores the capabilities of visual modalities in activity recognition. For example, Pose-on-the-Go~\cite{Ahuja2021PoseontheGoAU} combines RGB and IMU data at the sensor level to enhance real-time motion recognition. Similarly, Zhu et al.~\cite{Zhu2023IntegratingGA} demonstrate the integration of visual sensing and machine learning for robust activity detection in dynamic environments. These examples illustrate how vision-based pathways can independently enhance system responsiveness and accuracy, addressing variability in user behaviors while maintaining high performance.
}

\revise{
Another pathway, ``Location-Standard-Vision-Hybrid-APIs-Spatial Computing and Perception,'' highlights the integration of visual modalities with location-based data for cross-domain applications. GazePointAR~\cite{Lee2024GazePointARAC} combines visual and positional inputs within a hybrid framework to improve spatial computing accuracy, enabling seamless virtual-physical alignment in augmented reality environments. Zimmerer et al.~\cite{zimmerer2022case} further explore hybrid processing of LiDAR and RGB data for precise environmental mapping and navigation. These cases demonstrate the value of cross-modal integration, where visual data complements other contextual streams to address challenges like noise and latency. Designing scalable and robust integration methods is crucial to supporting diverse, multimodal applications.
}

\subsection{Usage and Implications}
\revise{
Our taxonomy, including the Sankey diagram and interactive website, serves as both a practical tool and a guiding framework for designing robust VMIs. By visualizing connections across dimensions and adopting a data modality-driven perspective, it enables customized strategies for addressing real-world challenges. For example, the connection between System Factors and Activity Detection highlights the importance of hybrid data stream integration, as demonstrated by MicroCam~\cite{hu2023microcam}, which fuses motion and visual data for surface detection. Similarly, RASSAR~\cite{Su2024RASSARRA} showcases how multimodal integration improves accessibility and safety evaluations. Together, these tools facilitate efficient resource allocation, scalable designs, and systematic solutions.
}

\revise{
The findings provide both macro and micro-level insights for designing context-aware systems. At the macro level, they identify key nodes, such as System Factors, and critical pathways shaping adaptive, scalable architectures. At the micro level, the categorization in Appendix~\ref{literature_detail} informs specific decisions, such as prioritizing adaptive algorithms to handle environmental variability or employing multimodal synchronization techniques. For example, EyeMU~\cite{meyer2022u} demonstrates how synchronized visual and motion data enhance gesture recognition, addressing user variability. Applications like GazePointAR~\cite{Lee2024GazePointARAC} reveal how VMIs align spatial computing with physical contexts, overcoming challenges like latency and noise. These insights emphasize the importance of robust synchronization, adaptive algorithms, and user-centric designs. Future work could explore the integration of GenAI and large language models to further enrich multimodal interactions for increasingly complex scenarios.
}

\section{Conclusion}
This research presents a taxonomy of VMIs aimed at enhancing context awareness. By synthesizing recent findings, it identifies key trends in multimodal data integration, system design, and context-aware applications. The taxonomy categorizes existing approaches across domains such as education, healthcare, accessibility, and gaming, with a focus on integrating visual modality with other inputs like audio, physiological signals, and motion. It also examines system design considerations, highlighting how VMIs process complex contextual information to improve user interactions. Open challenges, including real-time processing, data synchronization, and new interaction paradigms, are discussed to inform future research. This framework provides a foundation for developing adaptive, context-aware systems that support intuitive human-computer interactions in diverse applications.


\begin{acks}
\revisesecond{This work was supported by the Australian Research Council (Grant Nos. DP190102068, DP200102612, LP210200656) and the Core Industry IT Convergence Program (Grant No. 20016243), facilitated by the Ministry of Trade, Industry \& Energy (MOTIE, Korea).}
\end{acks}

\bibliographystyle{ACM-Reference-Format}
\bibliography{sample}


\clearpage 
\onecolumn 
\appendix

\begin{appendix}

\renewcommand{\thesection}{\Alph{section}}

\section{Appendix A: PRISMA-style Literature Selection}
\label{select_literature}

\begin{figure*}[htbp]
    \centering
\includegraphics[width=0.8\linewidth]{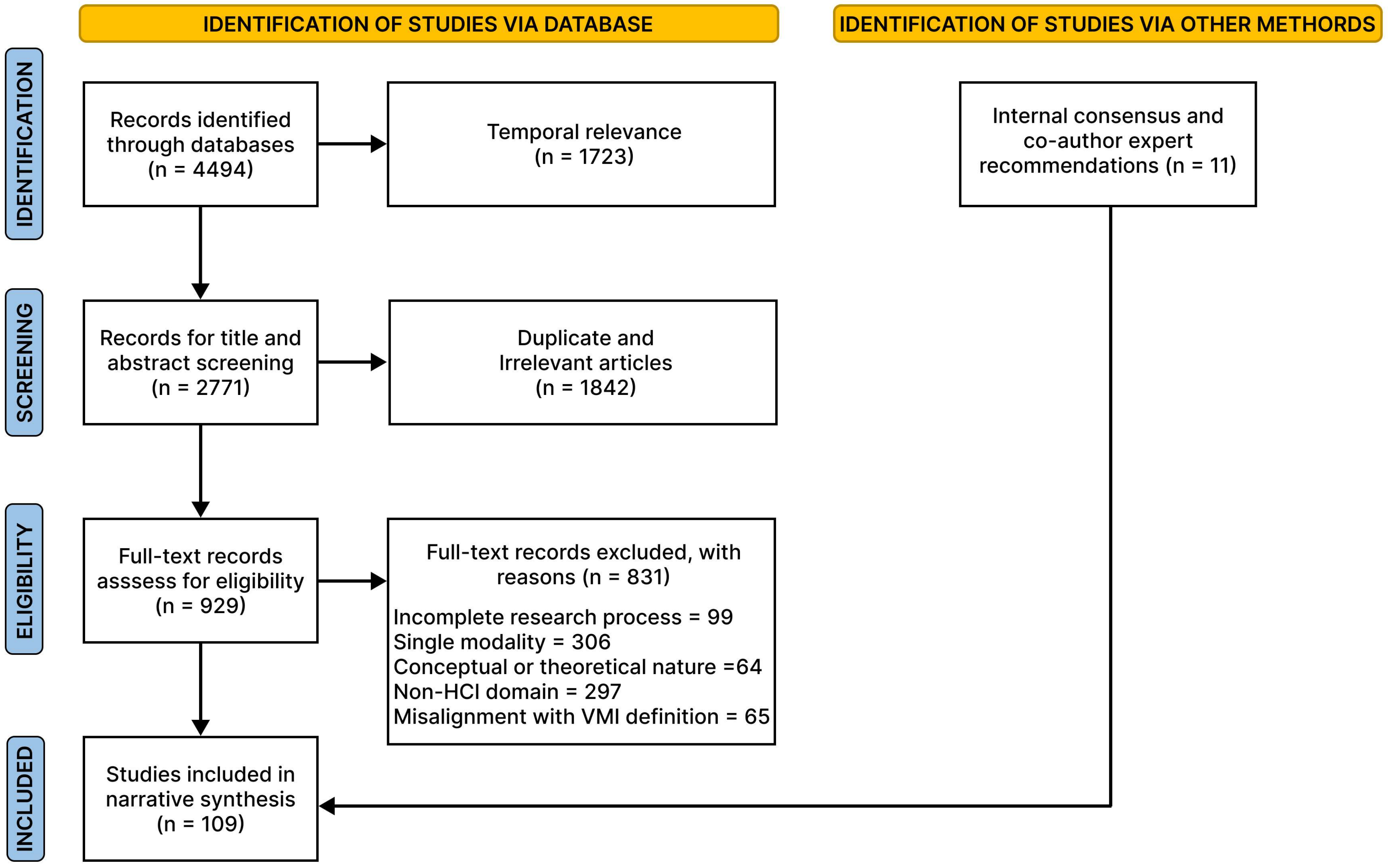}
\caption{A PRISMA-style flowchart of the selection of studies for the systematic review and meta-analysis.}
    \label{fig:PRISMA Framework}
\end{figure*}

Following the PRISMA framework~\cite{PRISMAStatement}, we conducted a four-stage systematic review, summarized in Figure~\ref{fig:PRISMA Framework}.

\textbf{Initial Phase:} \revisesecond{A keyword search was conducted in digital libraries, including ACM and IEEE databases, using \texttt{("vision-based") AND ("multimodal") AND ("context aware")} and related synonyms (e.g., \texttt{"vision"}, \texttt{"camera"}, \texttt{"visual"})}. This search yielded 4,494 works. A temporal filter was applied, excluding 1,723 items published in or before 2018.

\textbf{Screening Phase:} After removing duplicates and non-computational works, 1,842 exclusions were made. The remaining works were reviewed independently by two pairs of coders to ensure consistent application of criteria and inter-rater reliability~\cite{PRISMAStatement}. Coders assessed abstracts and full texts against predefined inclusion and exclusion criteria, and conflicts were resolved through the following steps:
\begin{itemize}
    \item Flagging conflicting cases for discussion in bi-weekly meetings.
    \item Applying a predefined resolution protocol, prioritizing alignment with the study scope.
    \item Revisiting discrepancies after each meeting to ensure a unified approach.
\end{itemize}

\textbf{Eligibility Phase:} Specific exclusion criteria were applied, resulting in the removal of 831 works and leaving 98 eligible papers. The excluded works fell into the following categories:
\begin{itemize}
    \item Incomplete research processes, such as workshops, symposia, tutorials, or technical briefs (99).
    \item Mentioning multimodality without utilizing it in the final system, or focusing solely on a single modality (306).
    \item Conceptual or theoretical works lacking demonstrable implementation, including case studies without prototypes or demos (64).
    \item Outside the HCI domain, unpublished in HCI-related venues (e.g., CHI, UIST), or lacking relevant keywords (297).
    \item Misaligned with the definition of VMIs, as described in Section~\ref{vmis_definition} (65).
\end{itemize}

\textbf{Final Phase:} Expert discussions added 11 relevant works to the dataset, resulting in a curated collection of 109 papers, primarily sourced from ACM (65\%), IEEE (23\%), and other libraries (12\%).

\revisesecond{
To ensure consistency and reduce subjectivity in coding, the following procedures were implemented across the phases:
\begin{itemize}
    \item Initial coding was performed on a small subset of the dataset to identify recurring themes and patterns.
    \item Discrepancies in coding were flagged and discussed during iterative group meetings, with consensus reached through majority agreement. In critical cases, an independent expert was consulted.
    \item Categories were refined through successive iterations, merging similar ones and expanding underrepresented dimensions.
    \item A final validation step was conducted on a randomly sampled subset (10\%) of the dataset to verify inter-coder reliability and alignment.
\end{itemize}
}

Through these steps, the final categorization of dimensions converged, ensuring robustness and consistency in the analysis process.

\clearpage 

\section{\revise{
Appendix B: Details of the Survey Literature Statistics} 
}
\label{literature_detail}

\begin{table*}[htbp]
\centering
\resizebox{\linewidth}{!}{
\begin{tabularx}{\textwidth}{p{0.3\textwidth} p{0.1\textwidth} p{0.6\textwidth}}
\vspace{0.4ex}Category & \vspace{0.4ex}Count & \vspace{0.4ex}Citations \\
\hline \\
\textbf{Section~\ref{taxonomy_context}: Context of VMIs} & &  \\
\textcolor{NavyBlue}{\textbf{\textit{Context Source Factors}}} & & \\
\vspace{0.4ex}\textcolor{NavyBlue}{\textit{$\rightarrow$ Human Factors}} & \vspace{0.4ex}87 &~\cite{Khan2021PALWA, Wang2024WatchYM, wang2024virtuwander, Su2024RASSARRA, li2024predicting, Ahuja2021PoseontheGoAU, Wang2024PepperPoseFP, Cai2024PANDALensTA, Wentzel2024SwitchSpaceUC, de2024llmr, Koch2023LeveragingDV, Saad2023HotFootFU, Lee2024GazePointARAC, Fan2024ContextCamBC, Liang2021AuthTrackEA, hoang2024artvista, Wen2024AdaptiveVoiceCA, schrapel2021spectrophone, Su2024SonifyARCS, Wen2024FindMT, Hautasaari2024EmoScribeCA, liao2024realityeffects, kong2021eyemu, Wang2023UbiPhysio, meyer2022u, zhang2024earsavas, Chen2021NeckFace, Zhu2023IntegratingGA, zhou2022gesture, Wang2024GVOILAGI, Kianpisheh2024exHAR, xu2024can, Meyer2021ACH, hu2023microcam, Fender2018VeltAF, shoa2023sushi, hu2024exploring, Liu2023MaTCRMT, Yang2023ContextAwareTV, Rachabatuni2024ContextawareCU, Yeo2017SpeCamSS, Gupta2023SensoryScapeCE, bokaris2019light, Wu2022ContextawareRD, fukasawa2022spatial, Bethge2021VEmotionUD, Dogan2021SensiCutML, cheng2021semanticadapt, Chen2023PaperToPlaceTI, Lian2023MultitaskLF, Li2021TowardsCA, hu2024multisurf, chen2024video2haptics, Fleer2012MISOAC, Matsuda2018EmoTourME, Liu2022VisuallyAwareAC, Wang2020CAPturARAA, Tsai2024GazeNoterCA, hu2023exploring, molina2023storytelling, Lin2024SelfCE, Xu2022MotionAC, Li2023TwoStepLS, Hallyburton2021SecurityAO, Deng2023ContextAwareFF, Malawade2022HydraFusionCS, Suzuki2019AnOM, Yookwan2022MultimodalFO, Bodi2024AIBasedES, Uimonen2023AGM, Schwrer2023Nav2CANAC, Zhao2022MediatedAT, Mittal2021MultimodalAC, Zheng2021StackedMA, Rawat2017ClickSmartAC, Jamonnak2021GeoContextAS, yin2024text2vrscene, Bang2021DeepLC, Niskanen2024EnhancingPC, Nouredanesh2022EgocentricVD, Harrison2008LightweightMD, kubo2017syncro, guarese2020augmented, somarathna2023exploring, marquardt2021airconstellations, wang2021elastic, shen2024mousering}
 \\
\vspace{0.4ex}\textcolor{NavyBlue}{\textit{$\rightarrow$ Environment Factors}}& \vspace{0.4ex}92 &~\cite{Wang2024WatchYM, wang2024virtuwander, Su2024RASSARRA, li2024predicting, Ahuja2021PoseontheGoAU, Wang2024PepperPoseFP, Cai2024PANDALensTA, Wentzel2024SwitchSpaceUC, de2024llmr, Koch2023LeveragingDV, Saad2023HotFootFU, Lee2024GazePointARAC, Lim2024ExploringCM, Cho2018DeepTI, Fan2024ContextCamBC, Liang2021AuthTrackEA, hoang2024artvista, Wen2024AdaptiveVoiceCA, ferdous2019s, zargham2024know, schrapel2021spectrophone, Su2024SonifyARCS, zimmerer2022case, kong2021eyemu, Wang2023UbiPhysio, meyer2022u, zhou2022gesture, Wang2024GVOILAGI, Kianpisheh2024exHAR, Meyer2021ACH, hu2023microcam, Fender2018VeltAF, hu2024exploring, Liu2023MaTCRMT, Yang2023ContextAwareTV, Rachabatuni2024ContextawareCU, Yeo2017SpeCamSS, Wu2022ContextawareRD, fukasawa2022spatial, Bethge2021VEmotionUD, Dogan2021SensiCutML, cheng2021semanticadapt, Hwang2020MonoEyeMH, Lian2023MultitaskLF, Li2021TowardsCA, hu2024multisurf, chen2024video2haptics, Fleer2012MISOAC, Liu2022VisuallyAwareAC, Wang2020CAPturARAA, Tsai2024GazeNoterCA, molina2023storytelling, Lin2024SelfCE, Xu2022MotionAC, McDuff2019AME, Li2023TwoStepLS, Liao2023GPT4EM, Hallyburton2021SecurityAO, Siddiqi2021AUA, Xie2023MFADAFUM, Deng2023ContextAwareFF, Malawade2022HydraFusionCS, su2021gaze+, Sun2023TemporallyCS, Yookwan2022MultimodalFO, Meng2022ValenceAA, Bodi2024AIBasedES, Uimonen2023AGM, Schwrer2023Nav2CANAC, Zhao2022MediatedAT, Mittal2021MultimodalAC, Shen2019ContextAwareDA, Athar2023VisTacTA, Zheng2021StackedMA, Deng2023AttentionAwareDN, Zhou2022ContextAware3O, Chou2020EncoderCameraGroundPR, Jamonnak2021GeoContextAS, yin2024text2vrscene, chen2024supporting, Bang2021DeepLC, Lin2014ACA, Cao2018DepthAT, Niskanen2024EnhancingPC, Nouredanesh2022EgocentricVD, Harrison2008LightweightMD, kubo2017syncro, guarese2020augmented, somarathna2023exploring, marquardt2021airconstellations, wang2021elastic, shen2024mousering}
\\
\vspace{0.4ex}\textcolor{NavyBlue}{\textit{$\rightarrow$ System Factors}} & \vspace{0.4ex}106 &~\cite{Khan2021PALWA, Wang2024WatchYM, wang2024virtuwander, Su2024RASSARRA, li2024predicting, Ahuja2021PoseontheGoAU, Wang2024PepperPoseFP, Cai2024PANDALensTA, Wentzel2024SwitchSpaceUC, Yang2012MagicFA, de2024llmr, Koch2023LeveragingDV, Saad2023HotFootFU, Lee2024GazePointARAC, Lim2024ExploringCM, Cho2018DeepTI, Fan2024ContextCamBC, Liang2021AuthTrackEA, hoang2024artvista, ferdous2019s, zargham2024know, schrapel2021spectrophone, Su2024SonifyARCS, Wen2024FindMT, Hautasaari2024EmoScribeCA, zimmerer2022case, liao2024realityeffects, kong2021eyemu, Wang2023UbiPhysio, meyer2022u, zhang2024earsavas, Chen2021NeckFace, Zhu2023IntegratingGA, zhou2022gesture, Wang2024GVOILAGI, Kianpisheh2024exHAR, xu2024can, Meyer2021ACH, hu2023microcam, Fender2018VeltAF, shoa2023sushi, hu2024exploring, Liu2023MaTCRMT, Yang2023ContextAwareTV, Rachabatuni2024ContextawareCU, Gupta2023SensoryScapeCE, bokaris2019light, Wu2022ContextawareRD, fukasawa2022spatial, Bethge2021VEmotionUD, Dogan2021SensiCutML, cheng2021semanticadapt, Chen2023PaperToPlaceTI, Hwang2020MonoEyeMH, Lian2023MultitaskLF, Li2021TowardsCA, hu2024multisurf, chen2024video2haptics, Fleer2012MISOAC, Matsuda2018EmoTourME, Liu2022VisuallyAwareAC, Wang2020CAPturARAA, Tsai2024GazeNoterCA, hu2023exploring, molina2023storytelling, Lin2024SelfCE, Xu2022MotionAC, McDuff2019AME, Li2023TwoStepLS, Liao2023GPT4EM, Hallyburton2021SecurityAO, Siddiqi2021AUA, Xie2023MFADAFUM, Deng2023ContextAwareFF, Malawade2022HydraFusionCS, Suzuki2019AnOM, su2021gaze+, Sun2023TemporallyCS, Yookwan2022MultimodalFO, Meng2022ValenceAA, Bodi2024AIBasedES, Uimonen2023AGM, Schwrer2023Nav2CANAC, Zhao2022MediatedAT, Mittal2021MultimodalAC, Shen2019ContextAwareDA, Athar2023VisTacTA, Rawat2017ClickSmartAC, Deng2023AttentionAwareDN, Zhou2022ContextAware3O, Chou2020EncoderCameraGroundPR, Jamonnak2021GeoContextAS, yin2024text2vrscene, chen2024supporting, Bang2021DeepLC, Lin2014ACA, Cao2018DepthAT, Niskanen2024EnhancingPC, Nouredanesh2022EgocentricVD, Harrison2008LightweightMD, kubo2017syncro, guarese2020augmented, somarathna2023exploring, marquardt2021airconstellations, wang2021elastic, shen2024mousering}
 \\
 \textcolor{NavyBlue}{\textbf{\textit{Context Categories}}} & & \\
\vspace{0.4ex}\textcolor{NavyBlue}{\textit{$\rightarrow$ Activity}} & \vspace{0.4ex}78
 &~\cite{Khan2021PALWA, Wang2024WatchYM, wang2024virtuwander, li2024predicting, Ahuja2021PoseontheGoAU, Wang2024PepperPoseFP, Cai2024PANDALensTA, Wentzel2024SwitchSpaceUC, Yang2012MagicFA, de2024llmr, Koch2023LeveragingDV, Lim2024ExploringCM, Liang2021AuthTrackEA, hoang2024artvista, Wen2024AdaptiveVoiceCA, ferdous2019s, zargham2024know, Su2024SonifyARCS, Wen2024FindMT, Hautasaari2024EmoScribeCA, zimmerer2022case, liao2024realityeffects, kong2021eyemu, Wang2023UbiPhysio, meyer2022u, zhang2024earsavas, Zhu2023IntegratingGA, zhou2022gesture, Wang2024GVOILAGI, Kianpisheh2024exHAR, xu2024can, Meyer2021ACH, hu2023microcam, Fender2018VeltAF, shoa2023sushi, hu2024exploring, Liu2023MaTCRMT, Yang2023ContextAwareTV, Rachabatuni2024ContextawareCU, bokaris2019light, Wu2022ContextawareRD, fukasawa2022spatial, Chen2023PaperToPlaceTI, Hwang2020MonoEyeMH, Li2021TowardsCA, hu2024multisurf, chen2024video2haptics, Fleer2012MISOAC, Matsuda2018EmoTourME, Liu2022VisuallyAwareAC, Wang2020CAPturARAA, Tsai2024GazeNoterCA, hu2023exploring, molina2023storytelling, Lin2024SelfCE, Xu2022MotionAC, Liao2023GPT4EM, Siddiqi2021AUA, Malawade2022HydraFusionCS, Suzuki2019AnOM, su2021gaze+, Uimonen2023AGM, Schwrer2023Nav2CANAC, Zhao2022MediatedAT, Mittal2021MultimodalAC, Athar2023VisTacTA, Zheng2021StackedMA, Rawat2017ClickSmartAC, Jamonnak2021GeoContextAS, chen2024supporting, Lin2014ACA, Nouredanesh2022EgocentricVD, Harrison2008LightweightMD, kubo2017syncro, somarathna2023exploring, marquardt2021airconstellations, wang2021elastic, shen2024mousering}
 \\
\vspace{0.4ex}\textcolor{NavyBlue}{\textit{$\rightarrow$ Location}}& \vspace{0.4ex}56 &~\cite{Khan2021PALWA, Wang2024WatchYM, wang2024virtuwander, Su2024RASSARRA, Cai2024PANDALensTA, Wentzel2024SwitchSpaceUC, Yang2012MagicFA, de2024llmr, Saad2023HotFootFU, Lee2024GazePointARAC, Lim2024ExploringCM, Cho2018DeepTI, Fan2024ContextCamBC, Wen2024AdaptiveVoiceCA, ferdous2019s, zargham2024know, schrapel2021spectrophone, Su2024SonifyARCS, zimmerer2022case, kong2021eyemu, Kianpisheh2024exHAR, xu2024can, hu2023microcam, Fender2018VeltAF, Yeo2017SpeCamSS, Wu2022ContextawareRD, fukasawa2022spatial, Bethge2021VEmotionUD, cheng2021semanticadapt, hu2024multisurf, chen2024video2haptics, Matsuda2018EmoTourME, Wang2020CAPturARAA, hu2023exploring, molina2023storytelling, Li2023TwoStepLS, Hallyburton2021SecurityAO, Xie2023MFADAFUM, Suzuki2019AnOM, Sun2023TemporallyCS, Yookwan2022MultimodalFO, Schwrer2023Nav2CANAC, Mittal2021MultimodalAC, Shen2019ContextAwareDA, Rawat2017ClickSmartAC, Zhou2022ContextAware3O, Chou2020EncoderCameraGroundPR, Jamonnak2021GeoContextAS, Bang2021DeepLC, Cao2018DepthAT, Niskanen2024EnhancingPC, Harrison2008LightweightMD, kubo2017syncro, guarese2020augmented, marquardt2021airconstellations, wang2021elastic}
 \\
\vspace{0.4ex}\textcolor{NavyBlue}{\textit{$\rightarrow$ Identity}} & \vspace{0.4ex}42
&~\cite{Khan2021PALWA, Su2024RASSARRA, Cai2024PANDALensTA, Wentzel2024SwitchSpaceUC, Yang2012MagicFA, de2024llmr, Cho2018DeepTI, Fan2024ContextCamBC, Liang2021AuthTrackEA, Wen2024AdaptiveVoiceCA, ferdous2019s, zargham2024know, schrapel2021spectrophone, Su2024SonifyARCS, Wen2024FindMT, Hautasaari2024EmoScribeCA, zimmerer2022case, zhang2024earsavas, Chen2021NeckFace, hu2023microcam, shoa2023sushi, Yeo2017SpeCamSS, Gupta2023SensoryScapeCE, bokaris2019light, Bethge2021VEmotionUD, Dogan2021SensiCutML, Lian2023MultitaskLF, hu2024multisurf, Matsuda2018EmoTourME, Lin2024SelfCE, McDuff2019AME, Liao2023GPT4EM, Deng2023ContextAwareFF, Meng2022ValenceAA, Bodi2024AIBasedES, Zhao2022MediatedAT, Mittal2021MultimodalAC, Deng2023AttentionAwareDN, Bang2021DeepLC, Lin2014ACA, Nouredanesh2022EgocentricVD, Harrison2008LightweightMD}
 \\
\vspace{0.4ex}\textcolor{NavyBlue}{\textit{$\rightarrow$ Time}} & \vspace{0.4ex}46 &~\cite{wang2024virtuwander, Su2024RASSARRA, li2024predicting, Cai2024PANDALensTA, Wentzel2024SwitchSpaceUC, de2024llmr, Lee2024GazePointARAC, Lim2024ExploringCM, Cho2018DeepTI, Liang2021AuthTrackEA, hoang2024artvista, ferdous2019s, zargham2024know, Su2024SonifyARCS, Wen2024FindMT, zimmerer2022case, liao2024realityeffects, kong2021eyemu, zhang2024earsavas, zhou2022gesture, Kianpisheh2024exHAR, shoa2023sushi, hu2024exploring, bokaris2019light, Wu2022ContextawareRD, cheng2021semanticadapt, hu2024multisurf, chen2024video2haptics, Matsuda2018EmoTourME, Wang2020CAPturARAA, hu2023exploring, molina2023storytelling, Xu2022MotionAC, su2021gaze+, Sun2023TemporallyCS, Meng2022ValenceAA, Rawat2017ClickSmartAC, Jamonnak2021GeoContextAS, yin2024text2vrscene, chen2024supporting, kubo2017syncro, guarese2020augmented, somarathna2023exploring, marquardt2021airconstellations, wang2021elastic, shen2024mousering}
 \\
\end{tabularx}
}
\end{table*}

\begin{table*}[htbp]
\centering
\renewcommand{\arraystretch}{1.5} 
\resizebox{\linewidth}{!}{
\begin{tabularx}{\textwidth}{p{0.3\textwidth} p{0.1\textwidth} p{0.6\textwidth}}
\vspace{0.4ex}Category & \vspace{0.4ex}Count & \vspace{0.4ex}Citations \\
\hline \\
\textbf{Section~\ref{sec: data modality}: Input Data Modality} & &  \\
 \textcolor{Plum}{\textbf{\textit{Visual Modality and Visual Dimensions}}} & & \\
\vspace{0.4ex}\textcolor{Plum}{\textit{$\rightarrow$ Standard-Vision}} & \vspace{0.4ex}108 &~\cite{Khan2021PALWA, Wang2024WatchYM, wang2024virtuwander, Su2024RASSARRA, li2024predicting, Ahuja2021PoseontheGoAU, Cai2024PANDALensTA, Wentzel2024SwitchSpaceUC, Yang2012MagicFA, de2024llmr, Koch2023LeveragingDV, Saad2023HotFootFU, Lee2024GazePointARAC, Lim2024ExploringCM, Cho2018DeepTI, Fan2024ContextCamBC, Liang2021AuthTrackEA, hoang2024artvista, Wen2024AdaptiveVoiceCA, ferdous2019s, zargham2024know, schrapel2021spectrophone, Su2024SonifyARCS, Wen2024FindMT, Hautasaari2024EmoScribeCA, zimmerer2022case, liao2024realityeffects, kong2021eyemu, Wang2023UbiPhysio, meyer2022u, zhang2024earsavas, Chen2021NeckFace, Zhu2023IntegratingGA, zhou2022gesture, Wang2024GVOILAGI, Kianpisheh2024exHAR, xu2024can, Meyer2021ACH, hu2023microcam, Fender2018VeltAF, shoa2023sushi, hu2024exploring, Liu2023MaTCRMT, Yang2023ContextAwareTV, Rachabatuni2024ContextawareCU, Yeo2017SpeCamSS, Gupta2023SensoryScapeCE, bokaris2019light, Wu2022ContextawareRD, fukasawa2022spatial, Bethge2021VEmotionUD, Dogan2021SensiCutML, cheng2021semanticadapt, Chen2023PaperToPlaceTI, Hwang2020MonoEyeMH, Lian2023MultitaskLF, Li2021TowardsCA, hu2024multisurf, chen2024video2haptics, Fleer2012MISOAC, Matsuda2018EmoTourME, Liu2022VisuallyAwareAC, Wang2020CAPturARAA, Tsai2024GazeNoterCA, hu2023exploring, molina2023storytelling, Lin2024SelfCE, Xu2022MotionAC, McDuff2019AME, Li2023TwoStepLS, Liao2023GPT4EM, Hallyburton2021SecurityAO, Siddiqi2021AUA, Xie2023MFADAFUM, Deng2023ContextAwareFF, Malawade2022HydraFusionCS, Suzuki2019AnOM, su2021gaze+, Sun2023TemporallyCS, Yookwan2022MultimodalFO, Meng2022ValenceAA, Bodi2024AIBasedES, Uimonen2023AGM, Schwrer2023Nav2CANAC, Zhao2022MediatedAT, Mittal2021MultimodalAC, Shen2019ContextAwareDA, Athar2023VisTacTA, Zheng2021StackedMA, Rawat2017ClickSmartAC, Deng2023AttentionAwareDN, Zhou2022ContextAware3O, Chou2020EncoderCameraGroundPR, Jamonnak2021GeoContextAS, yin2024text2vrscene, chen2024supporting, Bang2021DeepLC, Lin2014ACA, Cao2018DepthAT, Niskanen2024EnhancingPC, Nouredanesh2022EgocentricVD, Harrison2008LightweightMD, kubo2017syncro, guarese2020augmented, somarathna2023exploring, marquardt2021airconstellations, wang2021elastic, shen2024mousering}
\\
\vspace{0.4ex}\textcolor{Plum}{\textit{$\rightarrow$ Scale}}& \vspace{0.4ex}25 &~\cite{li2024predicting, Cai2024PANDALensTA, Yang2012MagicFA, Saad2023HotFootFU, Fan2024ContextCamBC, Liang2021AuthTrackEA, hoang2024artvista, schrapel2021spectrophone, hu2023microcam, Yeo2017SpeCamSS, Dogan2021SensiCutML, Hwang2020MonoEyeMH, hu2024multisurf, Tsai2024GazeNoterCA, Liao2023GPT4EM, Xie2023MFADAFUM, Deng2023ContextAwareFF, Malawade2022HydraFusionCS, Uimonen2023AGM, Zhao2022MediatedAT, Rawat2017ClickSmartAC, Zhou2022ContextAware3O, yin2024text2vrscene, Nouredanesh2022EgocentricVD, Harrison2008LightweightMD}
\\

\vspace{0.4ex}\textcolor{Plum}{\textit{$\rightarrow$ Spatial}} & \vspace{0.4ex}65 &~\cite{Khan2021PALWA, Wang2024WatchYM, wang2024virtuwander, li2024predicting, Ahuja2021PoseontheGoAU, Wang2024PepperPoseFP, Cai2024PANDALensTA, Wentzel2024SwitchSpaceUC, de2024llmr, Saad2023HotFootFU, Lee2024GazePointARAC, Fan2024ContextCamBC, Wen2024AdaptiveVoiceCA, ferdous2019s, Su2024SonifyARCS, Wen2024FindMT, zimmerer2022case, liao2024realityeffects, meyer2022u, Wang2024GVOILAGI, Fender2018VeltAF, shoa2023sushi, Gupta2023SensoryScapeCE, bokaris2019light, fukasawa2022spatial, Dogan2021SensiCutML, cheng2021semanticadapt, Chen2023PaperToPlaceTI, Hwang2020MonoEyeMH, Li2021TowardsCA, Fleer2012MISOAC, Matsuda2018EmoTourME, Wang2020CAPturARAA, Lin2024SelfCE, Xu2022MotionAC, Li2023TwoStepLS, Liao2023GPT4EM, Hallyburton2021SecurityAO, Siddiqi2021AUA, Xie2023MFADAFUM, Deng2023ContextAwareFF, Suzuki2019AnOM, Sun2023TemporallyCS, Yookwan2022MultimodalFO, Bodi2024AIBasedES, Uimonen2023AGM, Schwrer2023Nav2CANAC, Zhao2022MediatedAT, Mittal2021MultimodalAC, Shen2019ContextAwareDA, Athar2023VisTacTA, Zheng2021StackedMA, Rawat2017ClickSmartAC, Deng2023AttentionAwareDN, Zhou2022ContextAware3O, Chou2020EncoderCameraGroundPR, Jamonnak2021GeoContextAS, yin2024text2vrscene, chen2024supporting, Bang2021DeepLC, Lin2014ACA, Cao2018DepthAT, Nouredanesh2022EgocentricVD, guarese2020augmented, wang2021elastic}
\\
\vspace{0.4ex}\textcolor{Plum}{\textit{$\rightarrow$ Temporal}} & \vspace{0.4ex}47 &~\cite{Khan2021PALWA, Wang2024WatchYM, li2024predicting, Cai2024PANDALensTA, de2024llmr, Koch2023LeveragingDV, Saad2023HotFootFU, Lim2024ExploringCM, Liang2021AuthTrackEA, Wen2024AdaptiveVoiceCA, Su2024SonifyARCS, Wen2024FindMT, Hautasaari2024EmoScribeCA, zimmerer2022case, liao2024realityeffects, kong2021eyemu, zhang2024earsavas, Chen2021NeckFace, Zhu2023IntegratingGA, Wang2024GVOILAGI, Kianpisheh2024exHAR, Yang2023ContextAwareTV, Gupta2023SensoryScapeCE, Bethge2021VEmotionUD, Chen2023PaperToPlaceTI, Hwang2020MonoEyeMH, Li2021TowardsCA, chen2024video2haptics, Fleer2012MISOAC, Matsuda2018EmoTourME, Liu2022VisuallyAwareAC, Tsai2024GazeNoterCA, molina2023storytelling, Lin2024SelfCE, Xu2022MotionAC, Li2023TwoStepLS, Siddiqi2021AUA, Malawade2022HydraFusionCS, Sun2023TemporallyCS, Meng2022ValenceAA, Uimonen2023AGM, Zhao2022MediatedAT, Mittal2021MultimodalAC, Shen2019ContextAwareDA, Zheng2021StackedMA, Jamonnak2021GeoContextAS, Nouredanesh2022EgocentricVD}
 \\
\vspace{0.4ex}\textcolor{Plum}{\textit{$\rightarrow$ Beyond-Human-Vision}} & \vspace{0.4ex}22 &~\cite{Su2024RASSARRA, de2024llmr, Saad2023HotFootFU, Cho2018DeepTI, schrapel2021spectrophone, Chen2021NeckFace, Meyer2021ACH, Yeo2017SpeCamSS, fukasawa2022spatial, Lian2023MultitaskLF, hu2024multisurf, chen2024video2haptics, Li2023TwoStepLS, Hallyburton2021SecurityAO, Deng2023ContextAwareFF, Malawade2022HydraFusionCS, Bodi2024AIBasedES, Lin2014ACA, Cao2018DepthAT, Niskanen2024EnhancingPC, Harrison2008LightweightMD, shen2024mousering}
\\

\textcolor{Plum}{\textbf{\textit{Other Sensing Modalities}}} & & \\
\vspace{0.4ex}\textcolor{Plum}{\textit{$\rightarrow$ Audio}} & \vspace{0.4ex}36 &~\cite{Khan2021PALWA, wang2024virtuwander, Cai2024PANDALensTA, Lee2024GazePointARAC, Lim2024ExploringCM, hoang2024artvista, Wen2024AdaptiveVoiceCA, zargham2024know, Su2024SonifyARCS, Wen2024FindMT, Hautasaari2024EmoScribeCA, zimmerer2022case, liao2024realityeffects, kong2021eyemu, Wang2023UbiPhysio, meyer2022u, zhang2024earsavas, Wang2024GVOILAGI, xu2024can, Yang2023ContextAwareTV, Bethge2021VEmotionUD, Li2021TowardsCA, Fleer2012MISOAC, Matsuda2018EmoTourME, Liu2022VisuallyAwareAC, Tsai2024GazeNoterCA, Lin2024SelfCE, Xu2022MotionAC, McDuff2019AME, Malawade2022HydraFusionCS, su2021gaze+, Meng2022ValenceAA, Uimonen2023AGM, Zhao2022MediatedAT, Mittal2021MultimodalAC, Lin2014ACA}
 \\
\vspace{0.4ex}\textcolor{Plum}{\textit{$\rightarrow$ Text}}& \vspace{0.4ex}24 &~\cite{wang2024virtuwander, Cai2024PANDALensTA, de2024llmr, Lim2024ExploringCM, Fan2024ContextCamBC, hoang2024artvista, Su2024SonifyARCS, zhou2022gesture, Wang2024GVOILAGI, Kianpisheh2024exHAR, xu2024can, shoa2023sushi, hu2024exploring, Liu2023MaTCRMT, Rachabatuni2024ContextawareCU, Bethge2021VEmotionUD, hu2024multisurf, hu2023exploring, Liao2023GPT4EM, Meng2022ValenceAA, Mittal2021MultimodalAC, Zheng2021StackedMA, Rawat2017ClickSmartAC, yin2024text2vrscene}
 \\
\vspace{0.4ex}\textcolor{Plum}{\textit{$\rightarrow$ Motion}} & \vspace{0.4ex}34 &~\cite{Ahuja2021PoseontheGoAU, Wang2024PepperPoseFP, Cai2024PANDALensTA, de2024llmr, Koch2023LeveragingDV, Lim2024ExploringCM, Liang2021AuthTrackEA, ferdous2019s, Su2024SonifyARCS, kong2021eyemu, Wang2023UbiPhysio, meyer2022u, zhang2024earsavas, Zhu2023IntegratingGA, Kianpisheh2024exHAR, xu2024can, Meyer2021ACH, hu2023microcam, Wu2022ContextawareRD, fukasawa2022spatial, Bethge2021VEmotionUD, Fleer2012MISOAC, Matsuda2018EmoTourME, Wang2020CAPturARAA, Lin2024SelfCE, McDuff2019AME, Li2023TwoStepLS, Malawade2022HydraFusionCS, Uimonen2023AGM, Mittal2021MultimodalAC, Nouredanesh2022EgocentricVD, kubo2017syncro, somarathna2023exploring, shen2024mousering}
 \\
\vspace{0.4ex}\textcolor{Plum}{\textit{$\rightarrow$ Haptic}} & \vspace{0.4ex}3 &~\cite{Athar2023VisTacTA, Ahuja2021PoseontheGoAU, wang2021elastic}
 \\
\vspace{0.4ex}\textcolor{Plum}{\textit{$\rightarrow$ Positional}} & \vspace{0.4ex}9 &~\cite{Khan2021PALWA, Wen2024FindMT, Wu2022ContextawareRD, Bethge2021VEmotionUD, Malawade2022HydraFusionCS, Chou2020EncoderCameraGroundPR, Bang2021DeepLC, Niskanen2024EnhancingPC, marquardt2021airconstellations}
 \\
\vspace{0.4ex}\textcolor{Plum}{\textit{$\rightarrow$ Physiological}} & \vspace{0.4ex}8 &~\cite{Khan2021PALWA, Wen2024AdaptiveVoiceCA, hu2024exploring, Gupta2023SensoryScapeCE, Bethge2021VEmotionUD, Matsuda2018EmoTourME, Zhao2022MediatedAT, somarathna2023exploring}
 \\

\end{tabularx}
}
\end{table*}

\begin{table*}[htbp]
\centering
\renewcommand{\arraystretch}{1.5} 
\resizebox{\linewidth}{!}{
\begin{tabularx}{\textwidth}{p{0.3\textwidth} p{0.1\textwidth} p{0.6\textwidth}}
\vspace{0.4ex}Category & \vspace{0.4ex}Count & \vspace{0.4ex}Citations \\
\hline \\
\textbf{Section~\ref{sec: system_design_foundations}: System Design Foundations} & &  \\
\textcolor{Bittersweet}{\textbf{\textit{Data Integration Stages}}} & & \\
\vspace{0.4ex}\textcolor{Bittersweet}{\textit{$\rightarrow$ Sensor-Level Integration}}& \vspace{0.4ex}27 &~\cite{Wang2024WatchYM, Ahuja2021PoseontheGoAU, Wang2024PepperPoseFP, Lim2024ExploringCM, Liang2021AuthTrackEA, Zhu2023IntegratingGA, Kianpisheh2024exHAR, Meyer2021ACH, hu2023microcam, Fender2018VeltAF, Yeo2017SpeCamSS, bokaris2019light, fukasawa2022spatial, Hwang2020MonoEyeMH, Fleer2012MISOAC, Li2023TwoStepLS, Hallyburton2021SecurityAO, Bodi2024AIBasedES, Uimonen2023AGM, Chou2020EncoderCameraGroundPR, Lin2014ACA, Niskanen2024EnhancingPC, Harrison2008LightweightMD, kubo2017syncro, guarese2020augmented, somarathna2023exploring, marquardt2021airconstellations}
\\
\vspace{0.4ex}\textcolor{Bittersweet}{\textit{$\rightarrow$ Feature-Level Integration}} & \vspace{0.4ex}16 &~\cite{li2024predicting, Saad2023HotFootFU, Cho2018DeepTI, schrapel2021spectrophone, Hautasaari2024EmoScribeCA, meyer2022u, Liu2023MaTCRMT, Wu2022ContextawareRD, Dogan2021SensiCutML, Lian2023MultitaskLF, Suzuki2019AnOM, Yookwan2022MultimodalFO, Deng2023AttentionAwareDN, Nouredanesh2022EgocentricVD, wang2021elastic, shen2024mousering}
\\
\vspace{0.4ex}\textcolor{Bittersweet}{\textit{$\rightarrow$ Information-Level Integration}} & \vspace{0.4ex}12 &~\cite{wang2024virtuwander, Wentzel2024SwitchSpaceUC, de2024llmr, Wen2024AdaptiveVoiceCA, zargham2024know, liao2024realityeffects, Bethge2021VEmotionUD, cheng2021semanticadapt, molina2023storytelling, Siddiqi2021AUA, Rawat2017ClickSmartAC, Jamonnak2021GeoContextAS}
 \\
\vspace{0.4ex}\textcolor{Bittersweet}{\textit{$\rightarrow$ Hybrid Integration}} & \vspace{0.4ex}54 &~\cite{Khan2021PALWA, Su2024RASSARRA, Cai2024PANDALensTA, Yang2012MagicFA, Koch2023LeveragingDV, Lee2024GazePointARAC, Fan2024ContextCamBC, hoang2024artvista, ferdous2019s, Su2024SonifyARCS, Wen2024FindMT, zimmerer2022case, kong2021eyemu, Wang2023UbiPhysio, zhang2024earsavas, Chen2021NeckFace, zhou2022gesture, Wang2024GVOILAGI, xu2024can, shoa2023sushi, hu2024exploring, Yang2023ContextAwareTV, Rachabatuni2024ContextawareCU, Gupta2023SensoryScapeCE, Chen2023PaperToPlaceTI, Li2021TowardsCA, hu2024multisurf, chen2024video2haptics, Matsuda2018EmoTourME, Liu2022VisuallyAwareAC, Wang2020CAPturARAA, Tsai2024GazeNoterCA, hu2023exploring, Lin2024SelfCE, Xu2022MotionAC, McDuff2019AME, Liao2023GPT4EM, Xie2023MFADAFUM, Deng2023ContextAwareFF, Malawade2022HydraFusionCS, su2021gaze+, Sun2023TemporallyCS, Meng2022ValenceAA, Schwrer2023Nav2CANAC, Zhao2022MediatedAT, Mittal2021MultimodalAC, Shen2019ContextAwareDA, Athar2023VisTacTA, Zheng2021StackedMA, Zhou2022ContextAware3O, yin2024text2vrscene, chen2024supporting, Bang2021DeepLC, Cao2018DepthAT}
\\

\textcolor{Bittersweet}{\textbf{\textit{Multimodal Data Processing}}} & & \\
\vspace{0.4ex}\textcolor{Bittersweet}{\textit{$\rightarrow$ Rapid Solution Prototyping via Foundational Model APIs}} & \vspace{0.4ex}38 &~\cite{Wang2024WatchYM, wang2024virtuwander, Cai2024PANDALensTA, Wentzel2024SwitchSpaceUC, de2024llmr, Koch2023LeveragingDV, Lee2024GazePointARAC, Lim2024ExploringCM, Cho2018DeepTI, Fan2024ContextCamBC, Liang2021AuthTrackEA, hoang2024artvista, Wen2024FindMT, Hautasaari2024EmoScribeCA, zimmerer2022case, liao2024realityeffects, kong2021eyemu, Wang2023UbiPhysio, Wang2024GVOILAGI, Kianpisheh2024exHAR, xu2024can, shoa2023sushi, hu2024exploring, Rachabatuni2024ContextawareCU, Gupta2023SensoryScapeCE, cheng2021semanticadapt, Chen2023PaperToPlaceTI, hu2024multisurf, chen2024video2haptics, Wang2020CAPturARAA, Tsai2024GazeNoterCA, molina2023storytelling, McDuff2019AME, su2021gaze+, Schwrer2023Nav2CANAC, Zhao2022MediatedAT, Athar2023VisTacTA, yin2024text2vrscene}
\\
\vspace{0.4ex}\textcolor{Bittersweet}{\textit{$\rightarrow$ Developing A Dedicated ML Models}}& \vspace{0.4ex}91 &~\cite{Khan2021PALWA, Su2024RASSARRA, li2024predicting, Ahuja2021PoseontheGoAU, Wang2024PepperPoseFP, Wentzel2024SwitchSpaceUC, Yang2012MagicFA, de2024llmr, Koch2023LeveragingDV, Saad2023HotFootFU, Lee2024GazePointARAC, Lim2024ExploringCM, Cho2018DeepTI, hoang2024artvista, Wen2024AdaptiveVoiceCA, ferdous2019s, zargham2024know, schrapel2021spectrophone, Su2024SonifyARCS, Wen2024FindMT, Hautasaari2024EmoScribeCA, zimmerer2022case, liao2024realityeffects, kong2021eyemu, Wang2023UbiPhysio, meyer2022u, zhang2024earsavas, Chen2021NeckFace, Zhu2023IntegratingGA, zhou2022gesture, Wang2024GVOILAGI, Kianpisheh2024exHAR, xu2024can, Meyer2021ACH, hu2023microcam, hu2024exploring, Liu2023MaTCRMT, Yang2023ContextAwareTV, Yeo2017SpeCamSS, Gupta2023SensoryScapeCE, bokaris2019light, Wu2022ContextawareRD, fukasawa2022spatial, Bethge2021VEmotionUD, Dogan2021SensiCutML, cheng2021semanticadapt, Chen2023PaperToPlaceTI, Hwang2020MonoEyeMH, Lian2023MultitaskLF, Li2021TowardsCA, hu2024multisurf, chen2024video2haptics, Matsuda2018EmoTourME, Liu2022VisuallyAwareAC, Wang2020CAPturARAA, hu2023exploring, Lin2024SelfCE, Xu2022MotionAC, Li2023TwoStepLS, Liao2023GPT4EM, Siddiqi2021AUA, Xie2023MFADAFUM, Deng2023ContextAwareFF, Malawade2022HydraFusionCS, Suzuki2019AnOM, Sun2023TemporallyCS, Yookwan2022MultimodalFO, Meng2022ValenceAA, Bodi2024AIBasedES, Uimonen2023AGM, Mittal2021MultimodalAC, Shen2019ContextAwareDA, Athar2023VisTacTA, Zheng2021StackedMA, Deng2023AttentionAwareDN, Zhou2022ContextAware3O, Chou2020EncoderCameraGroundPR, Jamonnak2021GeoContextAS, yin2024text2vrscene, chen2024supporting, Bang2021DeepLC, Lin2014ACA, Cao2018DepthAT, Niskanen2024EnhancingPC, Nouredanesh2022EgocentricVD, kubo2017syncro, guarese2020augmented, somarathna2023exploring, marquardt2021airconstellations, wang2021elastic, shen2024mousering}
 \\
 
\vspace{0.4ex}\textcolor{Bittersweet}{\textit{$\rightarrow$ Heuristic Methods}} & \vspace{0.4ex}21 &~\cite{Yang2012MagicFA, Saad2023HotFootFU, zhang2024earsavas, hu2023microcam, Fender2018VeltAF, fukasawa2022spatial, Fleer2012MISOAC, Matsuda2018EmoTourME, Hallyburton2021SecurityAO, Siddiqi2021AUA, Suzuki2019AnOM, Meng2022ValenceAA, Uimonen2023AGM, Zhao2022MediatedAT, Mittal2021MultimodalAC, Rawat2017ClickSmartAC, Chou2020EncoderCameraGroundPR, chen2024supporting, Lin2014ACA, Harrison2008LightweightMD, somarathna2023exploring}
\\

\textcolor{Bittersweet}{\textbf{\textit{Evaluation Strategies}}} & & \\
\vspace{0.4ex}\textcolor{Bittersweet}{\textit{$\rightarrow$ Prototyping and Demonstration}} & \vspace{0.4ex}79 &~\cite{Khan2021PALWA, wang2024virtuwander, Su2024RASSARRA, Ahuja2021PoseontheGoAU, Wang2024PepperPoseFP, Cai2024PANDALensTA, Wentzel2024SwitchSpaceUC, Yang2012MagicFA, de2024llmr, Koch2023LeveragingDV, Saad2023HotFootFU, Lee2024GazePointARAC, Lim2024ExploringCM, Cho2018DeepTI, Fan2024ContextCamBC, Liang2021AuthTrackEA, hoang2024artvista, Wen2024AdaptiveVoiceCA, ferdous2019s, zargham2024know, Su2024SonifyARCS, Wen2024FindMT, Hautasaari2024EmoScribeCA, zimmerer2022case, liao2024realityeffects, kong2021eyemu, Wang2023UbiPhysio, Chen2021NeckFace, zhou2022gesture, Wang2024GVOILAGI, Kianpisheh2024exHAR, xu2024can, hu2023microcam, shoa2023sushi, hu2024exploring, Rachabatuni2024ContextawareCU, Gupta2023SensoryScapeCE, bokaris2019light, Wu2022ContextawareRD, fukasawa2022spatial, Dogan2021SensiCutML, cheng2021semanticadapt, Chen2023PaperToPlaceTI, Hwang2020MonoEyeMH, Li2021TowardsCA, hu2024multisurf, Fleer2012MISOAC, Wang2020CAPturARAA, Tsai2024GazeNoterCA, hu2023exploring, molina2023storytelling, McDuff2019AME, Hallyburton2021SecurityAO, Siddiqi2021AUA, Malawade2022HydraFusionCS, Suzuki2019AnOM, su2021gaze+, Meng2022ValenceAA, Bodi2024AIBasedES, Uimonen2023AGM, Schwrer2023Nav2CANAC, Zhao2022MediatedAT, Shen2019ContextAwareDA, Athar2023VisTacTA, Rawat2017ClickSmartAC, Chou2020EncoderCameraGroundPR, Jamonnak2021GeoContextAS, yin2024text2vrscene, chen2024supporting, Bang2021DeepLC, Lin2014ACA, Cao2018DepthAT, Nouredanesh2022EgocentricVD, Harrison2008LightweightMD, kubo2017syncro, guarese2020augmented, marquardt2021airconstellations, wang2021elastic, shen2024mousering}
\\

\vspace{0.4ex}\textcolor{Bittersweet}{\textit{$\rightarrow$ Technical Evaluation}}& \vspace{0.4ex}92 &~\cite{Wang2024WatchYM, Su2024RASSARRA, li2024predicting, Ahuja2021PoseontheGoAU, Wang2024PepperPoseFP, Wentzel2024SwitchSpaceUC, Yang2012MagicFA, de2024llmr, Koch2023LeveragingDV, Saad2023HotFootFU, Lee2024GazePointARAC, Cho2018DeepTI, Liang2021AuthTrackEA, Wen2024AdaptiveVoiceCA, schrapel2021spectrophone, Su2024SonifyARCS, Wen2024FindMT, kong2021eyemu, Wang2023UbiPhysio, meyer2022u, zhang2024earsavas, Chen2021NeckFace, Zhu2023IntegratingGA, zhou2022gesture, Wang2024GVOILAGI, Kianpisheh2024exHAR, xu2024can, Meyer2021ACH, hu2023microcam, Fender2018VeltAF, shoa2023sushi, hu2024exploring, Liu2023MaTCRMT, Yang2023ContextAwareTV, Rachabatuni2024ContextawareCU, Yeo2017SpeCamSS, bokaris2019light, fukasawa2022spatial, Bethge2021VEmotionUD, Dogan2021SensiCutML, Chen2023PaperToPlaceTI, Hwang2020MonoEyeMH, Lian2023MultitaskLF, Li2021TowardsCA, hu2024multisurf, chen2024video2haptics, Fleer2012MISOAC, Matsuda2018EmoTourME, Liu2022VisuallyAwareAC, Wang2020CAPturARAA, Tsai2024GazeNoterCA, Lin2024SelfCE, Xu2022MotionAC, McDuff2019AME, Li2023TwoStepLS, Liao2023GPT4EM, Hallyburton2021SecurityAO, Siddiqi2021AUA, Xie2023MFADAFUM, Deng2023ContextAwareFF, Malawade2022HydraFusionCS, Suzuki2019AnOM, su2021gaze+, Sun2023TemporallyCS, Yookwan2022MultimodalFO, Meng2022ValenceAA, Bodi2024AIBasedES, Uimonen2023AGM, Schwrer2023Nav2CANAC, Mittal2021MultimodalAC, Shen2019ContextAwareDA, Athar2023VisTacTA, Zheng2021StackedMA, Rawat2017ClickSmartAC, Deng2023AttentionAwareDN, Zhou2022ContextAware3O, Chou2020EncoderCameraGroundPR, Jamonnak2021GeoContextAS, yin2024text2vrscene, chen2024supporting, Bang2021DeepLC, Lin2014ACA, Cao2018DepthAT, Niskanen2024EnhancingPC, Nouredanesh2022EgocentricVD, Harrison2008LightweightMD, kubo2017syncro, guarese2020augmented, somarathna2023exploring, marquardt2021airconstellations, wang2021elastic, shen2024mousering}
 \\

\vspace{0.4ex}\textcolor{Bittersweet}{\textit{$\rightarrow$ User Evaluation}} & \vspace{0.4ex}76 &~\cite{Khan2021PALWA, Wang2024WatchYM, wang2024virtuwander, Su2024RASSARRA, li2024predicting, Ahuja2021PoseontheGoAU, Wang2024PepperPoseFP, Cai2024PANDALensTA, Wentzel2024SwitchSpaceUC, de2024llmr, Koch2023LeveragingDV, Lee2024GazePointARAC, Lim2024ExploringCM, Fan2024ContextCamBC, Liang2021AuthTrackEA, hoang2024artvista, Wen2024AdaptiveVoiceCA, ferdous2019s, zargham2024know, Su2024SonifyARCS, Wen2024FindMT, Hautasaari2024EmoScribeCA, zimmerer2022case, liao2024realityeffects, kong2021eyemu, Wang2023UbiPhysio, meyer2022u, zhang2024earsavas, Chen2021NeckFace, Zhu2023IntegratingGA, zhou2022gesture, Wang2024GVOILAGI, Kianpisheh2024exHAR, xu2024can, Meyer2021ACH, hu2023microcam, shoa2023sushi, Liu2023MaTCRMT, Yang2023ContextAwareTV, Yeo2017SpeCamSS, Gupta2023SensoryScapeCE, Wu2022ContextawareRD, fukasawa2022spatial, Bethge2021VEmotionUD, Dogan2021SensiCutML, cheng2021semanticadapt, Chen2023PaperToPlaceTI, Hwang2020MonoEyeMH, Li2021TowardsCA, chen2024video2haptics, Fleer2012MISOAC, Matsuda2018EmoTourME, Wang2020CAPturARAA, Tsai2024GazeNoterCA, hu2023exploring, molina2023storytelling, Lin2024SelfCE, Liao2023GPT4EM, Hallyburton2021SecurityAO, Deng2023ContextAwareFF, Malawade2022HydraFusionCS, Suzuki2019AnOM, su2021gaze+, Sun2023TemporallyCS, Uimonen2023AGM, Zhao2022MediatedAT, Rawat2017ClickSmartAC, Jamonnak2021GeoContextAS, yin2024text2vrscene, chen2024supporting, Harrison2008LightweightMD, guarese2020augmented, somarathna2023exploring, marquardt2021airconstellations, wang2021elastic, shen2024mousering}
 \\

\end{tabularx}
}
\end{table*}

\begin{table*}[htbp]
\centering
\renewcommand{\arraystretch}{1.5} 
\resizebox{\linewidth}{!}{
\begin{tabularx}{\textwidth}{p{0.3\textwidth} p{0.1\textwidth} p{0.6\textwidth}}
\vspace{0.4ex}Category & \vspace{0.4ex}Count & \vspace{0.4ex}Citations \\
\hline \\
\textbf{Section~\ref{sec: application_domains}: Application Domains} & &  \\
\textcolor{Dandelion}{\textit{$\rightarrow$ Location and Identity Recognition}} & 19 & \cite{Yang2012MagicFA, Saad2023HotFootFU, Lim2024ExploringCM, Liang2021AuthTrackEA, schrapel2021spectrophone, Wang2024GVOILAGI, xu2024can, hu2023microcam, Fender2018VeltAF, Yeo2017SpeCamSS, Dogan2021SensiCutML, hu2024multisurf, Li2023TwoStepLS, Schwrer2023Nav2CANAC, Deng2023AttentionAwareDN, Zhou2022ContextAware3O, Chou2020EncoderCameraGroundPR, Bang2021DeepLC, Harrison2008LightweightMD}
\\

\vspace{0.4ex}\textcolor{Dandelion}{\textit{$\rightarrow$ Activity Detection and Understanding}}& \vspace{0.4ex}46 &~\cite{Khan2021PALWA, Ahuja2021PoseontheGoAU, Wang2024PepperPoseFP, zargham2024know, Hautasaari2024EmoScribeCA, kong2021eyemu, Wang2023UbiPhysio, meyer2022u, zhang2024earsavas, Chen2021NeckFace, Zhu2023IntegratingGA, zhou2022gesture, Kianpisheh2024exHAR, Meyer2021ACH, hu2023microcam, shoa2023sushi, Yeo2017SpeCamSS, Gupta2023SensoryScapeCE, Wu2022ContextawareRD, fukasawa2022spatial, Bethge2021VEmotionUD, Hwang2020MonoEyeMH, Li2021TowardsCA, chen2024video2haptics, Fleer2012MISOAC, Matsuda2018EmoTourME, Wang2020CAPturARAA, Tsai2024GazeNoterCA, hu2023exploring, Lin2024SelfCE, Xu2022MotionAC, McDuff2019AME, Siddiqi2021AUA, su2021gaze+, Uimonen2023AGM, Schwrer2023Nav2CANAC, Athar2023VisTacTA, Zheng2021StackedMA, chen2024supporting, Lin2014ACA, Nouredanesh2022EgocentricVD, Harrison2008LightweightMD, kubo2017syncro, guarese2020augmented, somarathna2023exploring, shen2024mousering}
 \\

\vspace{0.4ex}\textcolor{Dandelion}{\textit{$\rightarrow$ Autonomous and Assistive Driving}} & \vspace{0.4ex}11 &~\cite{Koch2023LeveragingDV, Wen2024AdaptiveVoiceCA, Wu2022ContextawareRD, Bethge2021VEmotionUD, Liao2023GPT4EM, Hallyburton2021SecurityAO, Deng2023ContextAwareFF, Malawade2022HydraFusionCS, Schwrer2023Nav2CANAC, Zhou2022ContextAware3O, Jamonnak2021GeoContextAS}
\\

\vspace{0.4ex}\textcolor{Dandelion}{\textit{$\rightarrow$ Content Retrieval, Editing and Creation}} & \vspace{0.4ex}29 &~\cite{Wang2024WatchYM, Cai2024PANDALensTA, de2024llmr, Lee2024GazePointARAC, Fan2024ContextCamBC, hoang2024artvista, zimmerer2022case, liao2024realityeffects, Zhu2023IntegratingGA, zhou2022gesture, Wang2024GVOILAGI, xu2024can, Liu2023MaTCRMT, Yang2023ContextAwareTV, Rachabatuni2024ContextawareCU, Li2021TowardsCA, Liu2022VisuallyAwareAC, Tsai2024GazeNoterCA, hu2023exploring, molina2023storytelling, Xu2022MotionAC, Liao2023GPT4EM, su2021gaze+, Meng2022ValenceAA, Mittal2021MultimodalAC, Zheng2021StackedMA, Rawat2017ClickSmartAC, yin2024text2vrscene, chen2024supporting}
\\

\vspace{0.4ex}\textcolor{Dandelion}{\textit{$\rightarrow$ Spatial Computing and Perception}} & \vspace{0.4ex}48 &~\cite{wang2024virtuwander, li2024predicting, Ahuja2021PoseontheGoAU, Wentzel2024SwitchSpaceUC, de2024llmr, Lee2024GazePointARAC, ferdous2019s, zimmerer2022case, liao2024realityeffects, meyer2022u, Wang2024GVOILAGI, Fender2018VeltAF, shoa2023sushi, Gupta2023SensoryScapeCE, bokaris2019light, fukasawa2022spatial, Dogan2021SensiCutML, cheng2021semanticadapt, Chen2023PaperToPlaceTI, Hwang2020MonoEyeMH, Lian2023MultitaskLF, hu2024multisurf, chen2024video2haptics, Wang2020CAPturARAA, Tsai2024GazeNoterCA, hu2023exploring, molina2023storytelling, Li2023TwoStepLS, Deng2023ContextAwareFF, Suzuki2019AnOM, Sun2023TemporallyCS, Yookwan2022MultimodalFO, Bodi2024AIBasedES, Shen2019ContextAwareDA, Athar2023VisTacTA, Chou2020EncoderCameraGroundPR, chen2024supporting, Bang2021DeepLC, Lin2014ACA, Cao2018DepthAT, Niskanen2024EnhancingPC, Nouredanesh2022EgocentricVD, kubo2017syncro, guarese2020augmented, somarathna2023exploring, marquardt2021airconstellations, wang2021elastic, shen2024mousering}
\\

\vspace{0.4ex}\textcolor{Dandelion}{\textit{$\rightarrow$ Well-being and Health Care}} & \vspace{0.4ex}28 &~\cite{Khan2021PALWA, Wang2024WatchYM, Wang2024PepperPoseFP, Saad2023HotFootFU, Lim2024ExploringCM, Hautasaari2024EmoScribeCA, Wang2023UbiPhysio, meyer2022u, zhang2024earsavas, Kianpisheh2024exHAR, xu2024can, hu2024exploring, Gupta2023SensoryScapeCE, Matsuda2018EmoTourME, Lin2024SelfCE, Xu2022MotionAC, McDuff2019AME, Siddiqi2021AUA, Xie2023MFADAFUM, Meng2022ValenceAA, Zhao2022MediatedAT, Mittal2021MultimodalAC, Shen2019ContextAwareDA, Cao2018DepthAT, Niskanen2024EnhancingPC, Nouredanesh2022EgocentricVD, somarathna2023exploring, wang2021elastic}
 \\

\vspace{0.4ex}\textcolor{Dandelion}{\textit{$\rightarrow$ Education}} & \vspace{0.4ex}16 &~\cite{Wang2024PepperPoseFP, de2024llmr, ferdous2019s, Su2024SonifyARCS, Wang2023UbiPhysio, Zhu2023IntegratingGA, zhou2022gesture, Kianpisheh2024exHAR, shoa2023sushi, Rachabatuni2024ContextawareCU, Chen2023PaperToPlaceTI, Wang2020CAPturARAA, Tsai2024GazeNoterCA, molina2023storytelling, Zheng2021StackedMA, guarese2020augmented}
\\

\vspace{0.4ex}\textcolor{Dandelion}{\textit{$\rightarrow$ Accessibility}} & \vspace{0.4ex}27 &~\cite{Su2024RASSARRA, Yang2012MagicFA, Lee2024GazePointARAC, Cho2018DeepTI, hoang2024artvista, Wen2024FindMT, kong2021eyemu, meyer2022u, Fender2018VeltAF, Wu2022ContextawareRD, Dogan2021SensiCutML, Chen2023PaperToPlaceTI, Hwang2020MonoEyeMH, Lian2023MultitaskLF, hu2024multisurf, Fleer2012MISOAC, Liu2022VisuallyAwareAC, su2021gaze+, Uimonen2023AGM, Athar2023VisTacTA, chen2024supporting, Lin2014ACA, Nouredanesh2022EgocentricVD, kubo2017syncro, guarese2020augmented, wang2021elastic, shen2024mousering}
\\

\vspace{0.4ex}\textcolor{Dandelion}{\textit{$\rightarrow$ Game}} & \vspace{0.4ex}18 &~\cite{Ahuja2021PoseontheGoAU, chen2024video2haptics, de2024llmr, hu2023exploring, Hwang2020MonoEyeMH, Li2021TowardsCA, shen2024mousering, shoa2023sushi, somarathna2023exploring, Su2024SonifyARCS, Suzuki2019AnOM, Wang2024PepperPoseFP, Wang2020CAPturARAA, yin2024text2vrscene, zargham2024know, zhou2022gesture, zimmerer2022case, su2021gaze+}
 \\

\end{tabularx}
}
\end{table*}

\begin{table*}[htbp]
\centering
\renewcommand{\arraystretch}{1.5} 
\resizebox{\linewidth}{!}{
\begin{tabularx}{\textwidth}{p{0.3\textwidth} p{0.1\textwidth} p{0.6\textwidth}}
\vspace{0.4ex}Category & \vspace{0.4ex}Count & \vspace{0.4ex}Citations \\
\hline \\
\textbf{Section~\ref{sec: design_challenges}: Design Considerations and Key Challenges} & &  \\
\vspace{0.4ex}\textcolor{ForestGreen}{\textit{$\rightarrow$ Privacy and Security-aware Systems}} & \vspace{0.4ex}46 &~\cite{Khan2021PALWA, Wang2024WatchYM, Su2024RASSARRA, li2024predicting, Wang2024PepperPoseFP, Cai2024PANDALensTA, Wentzel2024SwitchSpaceUC, de2024llmr, Koch2023LeveragingDV, Saad2023HotFootFU, Lee2024GazePointARAC, Cho2018DeepTI, Fan2024ContextCamBC, Liang2021AuthTrackEA, zargham2024know, Hautasaari2024EmoScribeCA, meyer2022u, zhang2024earsavas, Zhu2023IntegratingGA, zhou2022gesture, Wang2024GVOILAGI, xu2024can, hu2023microcam, hu2024exploring, Rachabatuni2024ContextawareCU, Yeo2017SpeCamSS, Bethge2021VEmotionUD, Dogan2021SensiCutML, Hwang2020MonoEyeMH, hu2024multisurf, Liu2022VisuallyAwareAC, Tsai2024GazeNoterCA, McDuff2019AME, Liao2023GPT4EM, Hallyburton2021SecurityAO, Siddiqi2021AUA, Malawade2022HydraFusionCS, Uimonen2023AGM, Zhao2022MediatedAT, Shen2019ContextAwareDA, Jamonnak2021GeoContextAS, yin2024text2vrscene, chen2024supporting, Bang2021DeepLC, Harrison2008LightweightMD, shen2024mousering}
\\
\vspace{0.4ex}\textcolor{ForestGreen}{\textit{$\rightarrow$ User Variability}} & \vspace{0.4ex}51 &~\cite{Khan2021PALWA, Wang2024WatchYM, Su2024RASSARRA, Ahuja2021PoseontheGoAU, Wang2024PepperPoseFP, Wentzel2024SwitchSpaceUC, Yang2012MagicFA, Lee2024GazePointARAC, Lim2024ExploringCM, Fan2024ContextCamBC, Liang2021AuthTrackEA, hoang2024artvista, Wen2024AdaptiveVoiceCA, ferdous2019s, zargham2024know, Su2024SonifyARCS, Wen2024FindMT, Hautasaari2024EmoScribeCA, kong2021eyemu, Wang2023UbiPhysio, meyer2022u, zhang2024earsavas, Chen2021NeckFace, Zhu2023IntegratingGA, Wang2024GVOILAGI, Kianpisheh2024exHAR, xu2024can, Meyer2021ACH, hu2023microcam, shoa2023sushi, Yang2023ContextAwareTV, Rachabatuni2024ContextawareCU, Bethge2021VEmotionUD, cheng2021semanticadapt, Fleer2012MISOAC, Wang2020CAPturARAA, Tsai2024GazeNoterCA, hu2023exploring, molina2023storytelling, McDuff2019AME, Siddiqi2021AUA, Suzuki2019AnOM, Uimonen2023AGM, Schwrer2023Nav2CANAC, Zhao2022MediatedAT, Jamonnak2021GeoContextAS, Niskanen2024EnhancingPC, kubo2017syncro, somarathna2023exploring, marquardt2021airconstellations, shen2024mousering}
 \\

\vspace{0.4ex}\textcolor{ForestGreen}{\textit{$\rightarrow$ Ethics}} & \vspace{0.4ex}27 &~\cite{Khan2021PALWA, Wang2024WatchYM, Wang2024PepperPoseFP, Cai2024PANDALensTA, de2024llmr, Cho2018DeepTI, Hautasaari2024EmoScribeCA, zimmerer2022case, zhang2024earsavas, Zhu2023IntegratingGA, Wang2024GVOILAGI, xu2024can, hu2023microcam, shoa2023sushi, Rachabatuni2024ContextawareCU, Bethge2021VEmotionUD, Dogan2021SensiCutML, McDuff2019AME, Liao2023GPT4EM, Hallyburton2021SecurityAO, Siddiqi2021AUA, Malawade2022HydraFusionCS, Uimonen2023AGM, Deng2023AttentionAwareDN, Jamonnak2021GeoContextAS, yin2024text2vrscene, guarese2020augmented}
 \\
\vspace{0.4ex}\textcolor{ForestGreen}{\textit{$\rightarrow$ Cognitive Load and User Engagement}} & \vspace{0.4ex}58 &~\cite{Khan2021PALWA, wang2024virtuwander, Su2024RASSARRA, li2024predicting, Wang2024PepperPoseFP, Cai2024PANDALensTA, Wentzel2024SwitchSpaceUC, Yang2012MagicFA, Koch2023LeveragingDV, Lee2024GazePointARAC, Lim2024ExploringCM, hoang2024artvista, Wen2024AdaptiveVoiceCA, ferdous2019s, zargham2024know, Su2024SonifyARCS, Wen2024FindMT, Hautasaari2024EmoScribeCA, zimmerer2022case, liao2024realityeffects, Wang2023UbiPhysio, Zhu2023IntegratingGA, zhou2022gesture, Wang2024GVOILAGI, xu2024can, Meyer2021ACH, hu2023microcam, hu2024exploring, Liu2023MaTCRMT, Yang2023ContextAwareTV, bokaris2019light, Wu2022ContextawareRD, fukasawa2022spatial, Dogan2021SensiCutML, Chen2023PaperToPlaceTI, Li2021TowardsCA, hu2024multisurf, chen2024video2haptics, Fleer2012MISOAC, Liu2022VisuallyAwareAC, Wang2020CAPturARAA, Tsai2024GazeNoterCA, molina2023storytelling, Lin2024SelfCE, McDuff2019AME, Liao2023GPT4EM, Siddiqi2021AUA, Malawade2022HydraFusionCS, su2021gaze+, Uimonen2023AGM, Zhao2022MediatedAT, Mittal2021MultimodalAC, Athar2023VisTacTA, Zheng2021StackedMA, Rawat2017ClickSmartAC, Harrison2008LightweightMD, somarathna2023exploring, wang2021elastic}
\\
\vspace{0.4ex}\textcolor{ForestGreen}{\textit{$\rightarrow$ Automated Sensor Configuration}} & \vspace{0.4ex}32 &~\cite{Wang2024WatchYM, Ahuja2021PoseontheGoAU, Wang2024PepperPoseFP, Cai2024PANDALensTA, de2024llmr, Lim2024ExploringCM, zimmerer2022case, liao2024realityeffects, kong2021eyemu, meyer2022u, zhang2024earsavas, Zhu2023IntegratingGA, zhou2022gesture, Wang2024GVOILAGI, hu2023microcam, Fender2018VeltAF, hu2024multisurf, Tsai2024GazeNoterCA, McDuff2019AME, Li2023TwoStepLS, Liao2023GPT4EM, Hallyburton2021SecurityAO, Siddiqi2021AUA, Malawade2022HydraFusionCS, Bodi2024AIBasedES, Uimonen2023AGM, Zhao2022MediatedAT, Chou2020EncoderCameraGroundPR, yin2024text2vrscene, Niskanen2024EnhancingPC, Harrison2008LightweightMD, shen2024mousering}
\\
\vspace{0.4ex}\textcolor{ForestGreen}{\textit{$\rightarrow$ Context Discovery}} & \vspace{0.4ex}48 &~\cite{Khan2021PALWA, Su2024RASSARRA, li2024predicting, Wang2024PepperPoseFP, Cai2024PANDALensTA, Wentzel2024SwitchSpaceUC, de2024llmr, Lee2024GazePointARAC, Lim2024ExploringCM, Fan2024ContextCamBC, Wen2024AdaptiveVoiceCA, zargham2024know, schrapel2021spectrophone, Su2024SonifyARCS, zimmerer2022case, liao2024realityeffects, meyer2022u, zhang2024earsavas, Zhu2023IntegratingGA, Wang2024GVOILAGI, hu2023microcam, Yang2023ContextAwareTV, Wu2022ContextawareRD, fukasawa2022spatial, Bethge2021VEmotionUD, cheng2021semanticadapt, Chen2023PaperToPlaceTI, hu2024multisurf, Wang2020CAPturARAA, Tsai2024GazeNoterCA, molina2023storytelling, Xu2022MotionAC, McDuff2019AME, Li2023TwoStepLS, Liao2023GPT4EM, Hallyburton2021SecurityAO, Siddiqi2021AUA, Malawade2022HydraFusionCS, Suzuki2019AnOM, Bodi2024AIBasedES, Uimonen2023AGM, Schwrer2023Nav2CANAC, Zhao2022MediatedAT, Rawat2017ClickSmartAC, Jamonnak2021GeoContextAS, Harrison2008LightweightMD, kubo2017syncro, guarese2020augmented}
\\
\vspace{0.4ex}\textcolor{ForestGreen}{\textit{$\rightarrow$ Semantic Multimodal Data Integration}} & \vspace{0.4ex}54 &~\cite{wang2024virtuwander, Su2024RASSARRA, Wang2024PepperPoseFP, Cai2024PANDALensTA, de2024llmr, Lee2024GazePointARAC, Lim2024ExploringCM, Fan2024ContextCamBC, Wen2024AdaptiveVoiceCA, Su2024SonifyARCS, Wang2023UbiPhysio, meyer2022u, zhang2024earsavas, Zhu2023IntegratingGA, Wang2024GVOILAGI, xu2024can, Meyer2021ACH, hu2023microcam, Fender2018VeltAF, hu2024exploring, Liu2023MaTCRMT, Yang2023ContextAwareTV, Rachabatuni2024ContextawareCU, fukasawa2022spatial, Dogan2021SensiCutML, cheng2021semanticadapt, Chen2023PaperToPlaceTI, Lian2023MultitaskLF, Li2021TowardsCA, hu2024multisurf, Liu2022VisuallyAwareAC, Tsai2024GazeNoterCA, molina2023storytelling, Xu2022MotionAC, McDuff2019AME, Li2023TwoStepLS, Liao2023GPT4EM, Hallyburton2021SecurityAO, Siddiqi2021AUA, Deng2023ContextAwareFF, Malawade2022HydraFusionCS, Bodi2024AIBasedES, Uimonen2023AGM, Zhao2022MediatedAT, Mittal2021MultimodalAC, Athar2023VisTacTA, Chou2020EncoderCameraGroundPR, Jamonnak2021GeoContextAS, Lin2014ACA, Cao2018DepthAT, Niskanen2024EnhancingPC, Harrison2008LightweightMD, guarese2020augmented, wang2021elastic}
\\
\vspace{0.4ex}\textcolor{ForestGreen}{\textit{$\rightarrow$ Multimodal Contextual Reasoning}} & \vspace{0.4ex}73 &~\cite{Khan2021PALWA, Wang2024WatchYM, Su2024RASSARRA, li2024predicting, Wang2024PepperPoseFP, Cai2024PANDALensTA, de2024llmr, Koch2023LeveragingDV, Saad2023HotFootFU, Lee2024GazePointARAC, Lim2024ExploringCM, Cho2018DeepTI, Fan2024ContextCamBC, Wen2024AdaptiveVoiceCA, zargham2024know, Su2024SonifyARCS, Wen2024FindMT, liao2024realityeffects, Wang2023UbiPhysio, meyer2022u, zhang2024earsavas, Zhu2023IntegratingGA, Wang2024GVOILAGI, Kianpisheh2024exHAR, xu2024can, Meyer2021ACH, hu2023microcam, hu2024exploring, Liu2023MaTCRMT, Yang2023ContextAwareTV, Rachabatuni2024ContextawareCU, Yeo2017SpeCamSS, Gupta2023SensoryScapeCE, Wu2022ContextawareRD, Bethge2021VEmotionUD, Dogan2021SensiCutML, cheng2021semanticadapt, Chen2023PaperToPlaceTI, Lian2023MultitaskLF, Li2021TowardsCA, hu2024multisurf, Matsuda2018EmoTourME, Liu2022VisuallyAwareAC, Wang2020CAPturARAA, Tsai2024GazeNoterCA, Lin2024SelfCE, Xu2022MotionAC, McDuff2019AME, Liao2023GPT4EM, Hallyburton2021SecurityAO, Siddiqi2021AUA, Malawade2022HydraFusionCS, Suzuki2019AnOM, su2021gaze+, Sun2023TemporallyCS, Meng2022ValenceAA, Uimonen2023AGM, Schwrer2023Nav2CANAC, Zhao2022MediatedAT, Mittal2021MultimodalAC, Shen2019ContextAwareDA, Athar2023VisTacTA, Rawat2017ClickSmartAC, Deng2023AttentionAwareDN, Zhou2022ContextAware3O, Jamonnak2021GeoContextAS, Lin2014ACA, Cao2018DepthAT, Niskanen2024EnhancingPC, Nouredanesh2022EgocentricVD, kubo2017syncro, somarathna2023exploring, wang2021elastic}
\\

\vspace{0.4ex}\textcolor{ForestGreen}{\textit{$\rightarrow$ Imbalanced Data}} & \vspace{0.4ex}27 &~\cite{wang2024virtuwander, Su2024RASSARRA, Wang2024PepperPoseFP, de2024llmr, Koch2023LeveragingDV, Cho2018DeepTI, schrapel2021spectrophone, Wen2024FindMT, meyer2022u, zhang2024earsavas, Zhu2023IntegratingGA, zhou2022gesture, Kianpisheh2024exHAR, hu2023microcam, Yeo2017SpeCamSS, Lian2023MultitaskLF, Wang2020CAPturARAA, Lin2024SelfCE, Siddiqi2021AUA, Sun2023TemporallyCS, Meng2022ValenceAA, Deng2023AttentionAwareDN, Zhou2022ContextAware3O, Jamonnak2021GeoContextAS, yin2024text2vrscene, Bang2021DeepLC, somarathna2023exploring}
 \\
\vspace{0.4ex}\textcolor{ForestGreen}{\textit{$\rightarrow$ Assessment Heterogeneity}} & \vspace{0.4ex}11 &~\cite{wang2024virtuwander, Su2024RASSARRA, Wang2024PepperPoseFP, Matsuda2018EmoTourME, Sun2023TemporallyCS, Meng2022ValenceAA, Mittal2021MultimodalAC, Jamonnak2021GeoContextAS, yin2024text2vrscene, Bang2021DeepLC, Nouredanesh2022EgocentricVD}
\\
\vspace{0.4ex}\textcolor{ForestGreen}{\textit{$\rightarrow$ Scalable Architecture}} & \vspace{0.4ex}63 &~\cite{Wang2024WatchYM, Su2024RASSARRA, Ahuja2021PoseontheGoAU, Wang2024PepperPoseFP, Yang2012MagicFA, de2024llmr, Koch2023LeveragingDV, Saad2023HotFootFU, Lee2024GazePointARAC, Lim2024ExploringCM, Cho2018DeepTI, Fan2024ContextCamBC, Liang2021AuthTrackEA, hoang2024artvista, Wen2024AdaptiveVoiceCA, schrapel2021spectrophone, Su2024SonifyARCS, Hautasaari2024EmoScribeCA, zimmerer2022case, liao2024realityeffects, kong2021eyemu, Wang2023UbiPhysio, meyer2022u, zhang2024earsavas, Chen2021NeckFace, Zhu2023IntegratingGA, Wang2024GVOILAGI, Kianpisheh2024exHAR, xu2024can, Meyer2021ACH, hu2023microcam, Fender2018VeltAF, shoa2023sushi, Yeo2017SpeCamSS, Gupta2023SensoryScapeCE, bokaris2019light, Bethge2021VEmotionUD, Dogan2021SensiCutML, hu2024multisurf, chen2024video2haptics, Fleer2012MISOAC, Tsai2024GazeNoterCA, hu2023exploring, Xu2022MotionAC, McDuff2019AME, Li2023TwoStepLS, Liao2023GPT4EM, Hallyburton2021SecurityAO, Siddiqi2021AUA, Deng2023ContextAwareFF, Malawade2022HydraFusionCS, Yookwan2022MultimodalFO, Bodi2024AIBasedES, Uimonen2023AGM, Zhao2022MediatedAT, Shen2019ContextAwareDA, Zhou2022ContextAware3O, Chou2020EncoderCameraGroundPR, yin2024text2vrscene, Cao2018DepthAT, Niskanen2024EnhancingPC, marquardt2021airconstellations, shen2024mousering}
 \\
\vspace{0.4ex}\textcolor{ForestGreen}{\textit{$\rightarrow$ Power Consumption}} & \vspace{0.4ex}28 &~\cite{Wang2024WatchYM, Ahuja2021PoseontheGoAU, Wang2024PepperPoseFP, Yang2012MagicFA, kong2021eyemu, Wang2023UbiPhysio, meyer2022u, zhang2024earsavas, Chen2021NeckFace, Meyer2021ACH, Yeo2017SpeCamSS, Wu2022ContextawareRD, fukasawa2022spatial, Hwang2020MonoEyeMH, hu2024multisurf, chen2024video2haptics, Tsai2024GazeNoterCA, McDuff2019AME, Liao2023GPT4EM, Siddiqi2021AUA, Xie2023MFADAFUM, Malawade2022HydraFusionCS, su2021gaze+, Yookwan2022MultimodalFO, Uimonen2023AGM, Zhao2022MediatedAT, Athar2023VisTacTA, shen2024mousering}
\\

\end{tabularx}
}
\end{table*}
\end{appendix}

\end{document}

%% file: preamble.tex
\AtBeginDocument{%
  \providecommand\BibTeX{{%
    \normalfont B\kern-0.5em{\scshape i\kern-0.25em b}\kern-0.8em\TeX}}}

\copyrightyear{2025}
\acmYear{2025}
\setcopyright{acmlicensed}
\acmConference[CHI '25]{CHI Conference on Human Factors in Computing Systems}{April 26-May 1, 2025}{Yokohama, Japan}
\acmBooktitle{CHI Conference on Human Factors in Computing Systems (CHI '25), April 26-May 1, 2025, Yokohama, Japan}
\acmDOI{10.1145/3706598.3714161}
\acmISBN{979-8-4007-1394-1/25/04}

\usepackage{multirow}
\usepackage{array}

\usepackage{tabularx}
\usepackage{makecell}

\NewEnviron{reviseHighlight}{%
  \textcolor{blue}{\BODY}
}

\newif\ifCOMMENTS
\COMMENTSfalse

\newif\ifREVISESECOND   
\REVISESECONDfalse

\ifCOMMENTS

\newcommand{\revise}[1]{\begin{reviseHighlight}#1\end{reviseHighlight}} 

\newcommand{\reviseSubSectionTitle}[1]{\color{blue}{#1}}

\else

\newcommand{\reviseSubSectionTitle}[1]{#1}
\newcommand{\revise}[1]{#1} 

\fi

\ifREVISESECOND
\newcommand{\revisesecond}[1]{\textcolor{blue}{#1}}  
\else
\newcommand{\revisesecond}[1]{#1}   
\fi

\makeatletter
\def\Hline{
  \noalign{\ifnum0=`}\fi\hrule \@height 4.\arrayrulewidth \futurelet
   \reserved@a\@xhline}
\makeatother